# A Design Theory for AI-Assisted Software Development Derived from Christopher Alexander's Theory of Form

Chien-Tsun Chen — Teddysoft, Taipei, Taiwan (teddy@teddysoft.tw; corresponding author)
Yu Chin Cheng — Computer Science and Information Engineering, National Taipei University of Technology, Taipei, Taiwan (yccheng@ntut.edu.tw)

October 2026

## Abstract

In AI-assisted software development, code generated by large language models (LLMs) cannot be assumed to meet specified requirements. Conventional correctness-assurance measures such as reviews, testing, design by contract, and static analysis remain applicable. Verification, however, must keep pace with an unreliable generator. The open question is which of these measures a sufficient harness needs and what role each plays. This paper proposes a design theory for AI-assisted software development, derived from Christopher Alexander's theory of form rather than from prompting or agentic-programming techniques. It also provides a methodology for applying the theory in practice.

The theory builds on Alexander's view of design as achieving *fit* between a form, the software under development, and its context, the demands that software must satisfy. On this account, fit can be perceived only negatively, through the absence of identified *misfits*. We apply this view to AI coding by first making the development organization's tradition explicit, then deriving the misfits from that tradition and from the problem's classification. The harness then checks the generated software against those misfits.

The theory models the LLM as a *non-native vernacular builder*: a builder trained on many codebases but native to none, whose output tends to drift toward mainstream conventions rather than the local tradition. For such a builder, the theory recreates in explicit machinery what Alexander's *unselfconscious* process supplied through tradition: adherence is enforced rather than left to the builder's judgment. From Alexander's analysis, we derive obligations that must be discharged for the absence of detected misfits to justify a stated claim of fit. We discharge these obligations by engineering four pieces of machinery that Alexander's vernacular cultures did not need. First, the problem is made explicit through a representation based on Jackson's problem frames, and the solution-side tradition is made explicit as a *four-form pattern language*. Second, a stack of deterministic detectors makes misfits manifest. Third, a *fix loop* drives repair until the gates pass or the problem is escalated to a human. Fourth, a human-gated legislative circuit governs changes to the representations and detectors themselves. We call the resulting methodology *Misfit-Governed Development* (MGD). The builder cannot change this machinery on its own authority. Building and maintaining such machinery is the emerging practice of *harness engineering*.

The resulting process separates an autonomous inner loop, where the LLM iterates against the gates, from a human outer loop, where specifications are judged against the world and revised along with the tradition when they no longer fit. We call this the *dual-loop* process. The inner loop checks three families of relations by machine: the program and the tests must each conform to the specification, and the program must pass the tests. The outer loop carries the one judgment the machinery cannot make on its own authority: whether the specification fits the world. Together they form the *$S = P = T = W$ assurance model*, where S stands for specification, P for program, T for tests, and W for world; the equals sign is a mnemonic for these relations rather than a claim of literal equality. We report empirical evidence from building and rebuilding a Scrum system whose local tradition draws on domain-driven design, event sourcing, Clean Architecture, and Command Query Responsibility Segregation. Its four event-sourced aggregates are built from 64 problem-frame specifications and verified by approximately 1,300 generated tests under two infrastructure profiles and by 28 blocking gates, one of which applies 188 machine-checkable rules. This experience addresses the *generativity* dimension of Alexander's 1996 OOPSLA challenge to the software community. The *moral* dimension cannot be discharged by a gate: whether the specification still fits the world requires human judgment, which the theory locates in the outer loop as an institutional obligation rather than a detection problem.

# 1. Introduction

Large language models (LLMs) can generate plausible code at a rate that makes exhaustive human review increasingly costly. Worse, generated code can make an entire class of failures unusually difficult to notice through conventional code review. A misunderstood requirement sometimes leaves visible traces: the code may read awkwardly, violate local conventions, or fail to compile. Generated code weakens the evidential value of those traces: it can express the same misunderstanding as fluent implementation. It may compile and pass superficial inspection while silently dropping a field, inverting a guard condition, or defaulting to mainstream conventions learned during training. The practice known as vibe coding [1], in which generated code is accepted on the strength of its apparent plausibility, leaves the correctness question unanswered.

The methodology described in this paper takes the opposite approach. It treats the generator as *permanently* unreliable. This is a trust policy, not a prediction about future model capability. Prompt engineering and more capable models may improve the odds of success, but improved odds do not by themselves turn an LLM into a trustworthy translator from intent to code. Better odds are a property of the generator. They do not establish that any particular generated system fits its context. Therefore, the methodology withdraws trust from the generator and places it in a human-governed verification stack that surrounds the generator. Recent work has begun to treat this surrounding machinery as an engineering object in its own right, cataloging the harness's runtime responsibilities [2] and evolving harnesses automatically against observed failures [3]. The design problem then becomes the following: what must the verification stack contain, and what theory can tell us when it is sufficient for a stated claim of fit?

Our answer returns to Christopher Alexander's classical theory of form, set out in *Notes on the Synthesis of Form* [4], which preceded the pattern-language work later embraced by software engineering. There Alexander analyzed design as the achievement of fit between a *form* (the thing we shape) and its *context* (the part of the world that places demands on it). He observed that good fit can be perceived only negatively. *Misfits*, by contrast, can be identified individually, enumerated, and detected. This observation has a direct implication for AI coding. *To the extent that the relevant misfits can be enumerated and mechanically detected, synthesizing the form ceases to be a design problem and becomes a technical one.* Fit is operationalized, relative to the enumeration, as the absence of every enumerated misfit. Functional incorrectness is only one class of misfit: a system can meet its functional requirements and still misfit its context by violating architectural, contractual, or local-tradition demands. Synthesis then reduces to eliminating those misfits in turn. The strength of the resulting fit claim depends on how complete the enumeration is. In practice the enumeration is never finished, and not every recognized misfit yields a mechanical detector. The methodology therefore treats the enumeration itself as a governed, growing artifact: every escape either extends or corrects the enumeration or receives an explicit ruling (Sections 4.4 and 5).

Alexander further distinguished the *unselfconscious* process of vernacular building cultures from the *self-conscious* process of modern design. In the unselfconscious process, a builder without a global model of the problem achieves fit by following tradition and correcting misfits one at a time as reality makes them manifest. The self-conscious process cannot reliably achieve fit because the interacting misfits are too complex to treat as a whole, while the conceptual categories introduced to decompose them can distort the real misfit structure [4].

By the time Alexander addressed OOPSLA in 1996, software engineering had adopted his patterns as an engineering vocabulary. Alexander questioned whether their generative and moral substance had carried over [5]. He challenged the field to make software genuinely *generative*, to aim at coherent wholes, and to make life better.

This paper takes up that challenge by developing a design theory in the established sense of the term [6, 7]. The theory is prescriptive, grounded in kernel theories, and testable. This is not another prompting technique or agentic workflow. The paper makes five contributions:

1. **A design theory for AI coding, operationalized through Misfit-Governed Development.** We reframe AI code generation as Alexander's form–context problem. Misfit enumeration becomes the basis of verification. Machine-checked fit is the absence of every misfit in the mechanized enumeration. The central question is therefore how complete the enumeration is, rather than how capable the generator is (Sections 2, 3, 4.2, 4.5, and 6; Table 1; Appendix A.1).
2. **The non-native vernacular builder.** We model the generator as a craftsman trained across many traditions and native to none, whose deviations pull persistently toward mainstream idiom. This model and software's silent misfits yield enforced tradition, manufactured manifestation, and severed backflow (Sections 3.5, 4.1, 4.4, and 4.5).
3. **The four-form pattern language.** The language's reader is a machine rather than a judicious professional. Every pattern node must exist in four coupled forms (doctrine, template, prohibition, gate), and the language's grammar is enforced rather than followed. It is paired with a problem representation whose frame catalog extends Jackson's. Each problem-frame type's workflow is an ordered traversal of the solution-side pattern network (Section 3).
4. **The dual-loop process and the S = P = T = W assurance model.** Building the machinery that surrounds the generator is a practice that has developed faster than its theory. We derive a theory for its verification core. The separation of machine-paced and human-paced feedback follows from the asymmetry between machine-priced and human-priced detection rather than from a design preference; the two loops are its realization. The resulting *dual-loop* process is the engineered analogue of Alexander's self-correcting vernacular process, yielding the *S = P = T = W assurance model*. Here, S denotes specification, P program, T tests, and W world. The model comprises three families of machine-decidable relations enclosed by a fourth relation, S = W, that requires human judgment; the equals signs abbreviate these relations, and Appendix A gives their formal semantics (Sections 3.1, 3.5–3.7, and 4.2–4.3).

5. **A graded answer to Alexander's OOPSLA challenge.** The methodology operationalizes generativity for software pattern languages. The LLM can interpret a pattern language as written while being deterministically governed by it. Coherence is answered in the formal sense (the parts verifiably compose) and left open in the experiential sense (Alexander's felt wholeness). The moral dimension, whether the built thing serves human life, enters the assurance model through the S = W relation, which no gate can discharge. It must therefore be answered by institutions rather than detectors (Section 6).

These contributions form a single chain rather than five independent claims. The builder model motivates the pattern language, the pattern language requires the harness that enforces it, and the asymmetry between machine-priced and human-priced detection separates that harness's feedback into the two loops of Contribution 4. Together they constitute the design theory introduced in Contribution 1 and ground its graded answer to Alexander's challenge (Contribution 5).

Section 2 reviews the theoretical background and states the derivation. Section 3 describes the methodology and its reference implementation. Section 4 develops the theoretical analysis. Section 5 reports the empirical evidence. Section 6 returns to Alexander's challenge. Sections 7–9 discuss related work, limitations, and conclusions.

# 2. Background and the Derivation

## 2.1 Fit, misfit, and the negative definition of good design

Alexander's *Notes on the Synthesis of Form* [4] frames every design problem as an *ensemble*: a form to be shaped and a context that places demands on it. Its central observation is an asymmetry of perception. Designers cannot perceive good fit positively. They can only detect *misfits* (specific, enumerable failures of the form to meet a demand). Fit is therefore assessed negatively, through the absence of identified misfits. We operationalize fit, relative to an enumerated misfit list, as the absence of every misfit on that list (Appendix A). How complete the list is becomes the central question. To the extent that the relevant misfits have been enumerated and made decidable, the verdict on a candidate form becomes mechanical. What lies outside that enumeration remains a matter for judgment. Alexander himself conceded that the *selection* of misfit variables cannot be formalized, a concession that, as we show in Sections 5 and 6 (see also Table 1, P8), predicts the residue of judgment that remains in our methodology.

Design research later re-examined this negative formulation. Salustri [8] judges *Notes* "flawed, but not flawed beyond use": he keeps the misfit core, but because a representation that records only unsatisfactory aspects can miss a repair's effect on aspects that were previously satisfactory, he extends it into a balance among interacting properties. We retain the negative formulation for verification and instead make its scope explicit: fit is claimed only relative to the enumerated misfits (Appendix A). Every repair re-runs the whole enumeration, so a disturbed aspect that the enumeration covers appears as a new detected misfit (Section 3.6), while misfits outside the enumeration remain the outer loop's responsibility (Section 4.4).

## 2.2 The unselfconscious process and its enabling conditions

Alexander's account of vernacular building cultures [4] describes a process that achieves fit without any participant holding a global model. The builder follows *tradition*, a crystallized memory of resolved misfits that includes taboos. Tradition also supplies *viscosity*: it resists change, so each generation of builders starts near known-good form. Reality makes misfits *manifest* (the roof leaks). Correction is then local and immediate. We emphasize three enabling conditions usually left implicit:

1. *Manifestation is physical*: failures convert themselves into sensory signals without engineered detectors.
2. *Perception is collocated*: those who suffer a misfit, those who notice it, and those who can repair it live with the building, often as one and the same person, so detection is a free by-product of dwelling.
3. *Interpretation is cultural*: tradition has pre-decided what counts as a misfit. The same draft is a defect in one building culture and desirable ventilation in another.

Under these conditions the process is *homeostatic* [4]. Good fit is maintained because rapid local correction closes the feedback loop between reality and form, and because the vernacular's misfit network decomposes into weakly coupled subsystems, so that repairs rarely propagate.

When Alexander later confronted a setting where these conditions fail (a modern university campus), he did not pretend they still held. He engineered a prosthetic mechanism. The Oregon Experiment's *diagnosis* principle [9] institutionalizes misfit perception. A periodic survey of the campus, scored against the pattern language, feeds a repair budget. This precedent matters because none of these conditions can be assumed to hold for the non-native software builder of this paper. Each will have to be engineered back (Section 3).

## 2.3 The arc to *The Timeless Way* and the OOPSLA challenge

The Oregon prosthetic repaired one institution's failing process from the outside. Alexander's larger project was to rebuild tradition itself. Between *Notes* and *The Timeless Way of Building* [10], his position matured into what we read as a three-stage arc. The shared, implicit languages of vernacular cultures have broken down. Self-conscious design fails by substituting the designer's arbitrary categories for the real misfit structure. The remedy is to *reconstruct tradition self-consciously* through an explicit pattern language whose patterns state the forces they resolve [11], with the ultimate aim of re-internalization. The final chapter of *The Timeless Way* treats the explicit language as scaffolding: the master builder "no longer needs the language" and responds directly to the forces themselves [10]. The image is Alexander's own: the part of the book that presents the pattern language is titled "The Gate," and the way is not complete "until we leave the gate behind" [10]. In this account, the explicit language is a gate that the builder passes through and eventually leaves behind, a passage toward mastery rather than a permanent checkpoint. We call this hoped-for endpoint, in which the builder acts from the tradition without consulting the language, Alexander's *forgetting*.

At OOPSLA 1996 [5], Alexander told the software patterns community that it had taken his vocabulary, and he questioned whether three central concerns had come with it: the *moral* dimension, *coherence*, and *generativity*. The moral dimension asks whether patterns make life better, coherence whether the built result hangs together as a single whole, and generativity whether a language generates coherent wholes the way a genome generates an organism rather than merely serving as a catalog of parts. He closed by asking whether software (a field with more practitioners and more leverage than architecture) would use its computational power to build genuinely generative processes.

## 2.4 Problem frames

While Alexander supplies the theory of fit, Jackson supplies the vocabulary that we use to state the problem. Michael Jackson's problem frames [12] are, in our reading, the software analogue of Alexander's ensemble. *The problem is in the world; the machine is the solution*. Analysis consists of drawing the boundary between them and classifying the problem into recurring frame types, each carrying a known set of *frame concerns*. In Alexandrian terms, that set supplies a pre-enumerated core of the misfit list whose cost is amortized across every problem of that class. This amortization makes broad enumeration operational for the recurring part of the context. No project can enumerate its context from scratch, but a project that classifies its problems into frames inherits a reusable starting structure for the enumeration. The project must still enumerate its own domain-specific misfits on top of that structure (Section 3.2).

## 2.5 From propositions to obligations: the derivation, stated explicitly

The title of this paper claims that our theory is derived from Alexander's theory of form, not merely inspired by it, and this claim should be supported explicitly. Table 1 states the mapping. Each row names a proposition of the classical theory (stated explicitly by Alexander, in P7 and P9 drawing on his later work [9, 10], or reconstructed from the mechanisms he describes), the engineering obligation it imposes once the builder is the non-native LLM of Section 4.1, and the mechanism that discharges it. None of the entries is arbitrary. The derivation fixes the obligations but not the mechanisms. Each obligation follows from the proposition in the left-hand column under the engineering setting considered here, together with, where relevant, the premise that the builder is a non-native LLM. The mechanisms in the right-hand column are engineering choices. Each is one way of discharging its obligation but not the only way. The rows are presented separately for exposition, but they are not claimed to be logically independent: an obligation may draw on its proposition together with the builder premise and with obligations established by earlier rows. Alexander's propositions, the traditions the methodology draws on, and the S = P = T = W model were all in hand before the harness was built, and they guided its construction. The derivation, however, does not rest on that history. It stands or falls on whether each obligation follows from its stated proposition together with the builder premise and any prior obligations on which the row depends.

The ten rows are not claimed to be exhaustive. A proposition earns a row by one test: it must yield an engineering obligation once the builder premise changes. A proposition that yields no new obligation under that change is omitted. The best-known operational content of *Notes*, its formal procedure for decomposing misfit variables into hierarchical subsets, does not qualify for a row. It is a mechanism rather than a proposition, and its structural role is filled here by domain-driven decomposition (P5). This exclusion is also consistent with Alexander's own later assessment. In the preface to the paperback edition [4], he set the mathematics aside while retaining the account of fit and misfit on which the derivation builds.

Like the misfit list itself, the table remains open, but at a different level. An escape (a defect that the gates missed) is first answered within the existing rows: a new detector, a repaired one, or a reconciled proxy discharges the obligations already stated (Section 4.4; exercised in Section 5). Only an escape that exposes an engineering obligation not derivable from any existing row calls for a new row. Appendix A states the underlying misfit calculus formally.

The derivation claims unity, not exclusivity. Some obligations can also be justified by other software engineering theories. For example, the obligation that fit must rest on the verdict of detectors (P1) and the obligation to manufacture manifestation where misfits do not reveal themselves (P2) are consistent with long-standing verification practice, and the cadence asymmetry of P7 is consistent with iterative development. The claim is not that Alexander alone can justify these obligations. The claim is that all ten arise from a single account of fit rather than being assembled from unrelated engineering principles. Three of them also arrive with a different status from their nearest counterparts in current practice. Practice knows change control and verification independence as precautions. Here, the severed backflow between builder and tradition (P6), the permanence of the scaffold (P9), and the obligation to place every governing representation, not only the tests, under verification (P10) follow as consequences of the builder premise, from Alexander's analyses of vernacular culture, of internalization, and of representations. Sections 4.4, 4.5, and 4.2 develop them at length, and Observation 2 exercises P10 in operation. Table 1 should therefore be read as a derivation from one kernel theory, not as a claim that it is the only possible derivation. The division runs as follows. Five of the ten (P1, P2, P4, P5, P7) express requirements that do not depend specifically on the builder being an LLM, and a reader who already practices verification will recognize them. Four (P3, P6, P9, P10) take on their present force because the builder is an LLM: it cannot be shown to have internalized the local tradition, it defaults to the mainstream, and its access to context is mediated through representations. P8 sits between the two groups. The residue of judgment exists for any builder, but once the builder is mechanized, no one can establish that it carries that residue, so the obligation to place the residue outside the autonomous inner loop is specific to this setting. We state the division rather than leave it implicit, because the contribution is the derivation of all ten from one account of fit, not a claim that each is new.

*Table 1. The derivation: propositions of Alexander's classical theory, the obligations they impose once the builder is a non-native LLM, and the mechanisms that discharge them.*

| # | Proposition of the classical theory | Obligation once the builder is a non-native LLM | Mechanism (section) |
|---|---|---|---|
| P1 *negative fit* | Fit is perceived negatively, through identifiable misfits [4] | Machine-checked fit must be defined relative to the mechanized misfit enumeration and determined by misfit detectors, never by generator confidence | Gate stack; 188-rule catalog (§3.5) |
| P2 *manifestation* | In the vernacular, misfits manifest physically, without engineered detectors [4] | Software misfits do not reliably reveal themselves; manifestation must be manufactured | Deterministic gates, with fixtures showing that they still fire and real invocations showing that they are still reached (§3.4–3.5, §4.2) |
| P3 *viscosity* | Tradition supplies viscosity, so the builder starts near known-good form [4] | The builder cannot be assumed to carry the local tradition internally; the viscosity that tradition supplies must therefore be recreated outside the builder and enforced | Four-form pattern language over the architectural network, traversed per problem-frame type (§3.2–3.4) |
| P4 *stable context* | One-at-a-time correction converges only under a stable context [4] | Context must be frozen for the duration of each repair | Specification and framework version pinned inside the fix-loop window (§3.6) |

| # | Proposition of the classical theory | Obligation once the builder is a non-native LLM | Mechanism (section) |
|---|---|---|---|
| P5 *repair locality* | Local correction converges only if the effects of repair stay sufficiently local [4] | Repair propagation must be structurally bounded, and residual propagation must be detected | Aggregates coupled through events bound horizontal propagation, and Clean Architecture's inward-only dependencies bound vertical propagation (§3.3); the regression guard catches residual propagation (§3.6) |
| P6 *tradition-bearer* | In the vernacular, the builder is also a bearer of the tradition that governs building [4] | The authority to amend the tradition must be withheld from the builder; otherwise its learned defaults would pull the tradition toward them | Severed backflow: the generator may not amend the tradition; amendments pass through the human legislature (§3.7, §4.4) |
| P7 *detection cost* | Detection cost sets the affordable feedback cadence, read from the vernacular's embedded detection and Oregon's periodic diagnosis [4, 9] | Feedback must be scheduled at the cadence each detector class affords: continuous for machine-priced detection, budgeted for human-priced detection | Dual-loop architecture (§3.1, §4.3) |
| P8 *residue of judgment* | The selection of misfit variables cannot itself be formalized [4] | With the builder mechanized, the residue of judgment can no longer be presumed to reside in the builder; it must be explicitly placed outside the autonomous inner loop | S = W reserved to the human outer loop; the misfit list grown and amended only by the human legislature (§3.7, §5, §6) |
| P9 *scaffolding* | The explicit language is scaffolding: mastery is reached when the builder no longer needs the language and the forces are answered directly [10] | Observed compliance cannot justify retiring enforcement: machine-checked fit must continue to rest on detector verdicts, never on presumed internalization (P1) | Gates as permanent checkpoints; anti-mainstream rules never retire on statistical grounds; every repair re-anchored in the tradition (§3.6, §4.1, §4.5) |
| P10 *formal picture* | The designer works from a picture of the context, not the context itself; the remedy is to make the picture explicit, criticizable, and correctable [4] | The builder's access to context is mediated by representations, and representations drift silently; the proxies themselves must therefore be verified | Primary/proxy split with mechanized reconciliation: session-start checks, dual-profile execution, self-audit of the verification artifacts (§3.5, §4.2, §5) |

The table states the derivation proposition by proposition. Figure 1 condenses those obligations into the machinery of the next section. At design time (once, then grown by governance), Alexander's negative definition of fit is instantiated in three coupled artifacts: a problem representation (frame types with their concerns), a solution-side language (the pattern network of Section 3.3), and an enumerated misfit list (rules with their fixtures). At run time (per use case), a specification is interpreted by the LLM, and candidate forms iterate against the gate stack until no misfit manifests.

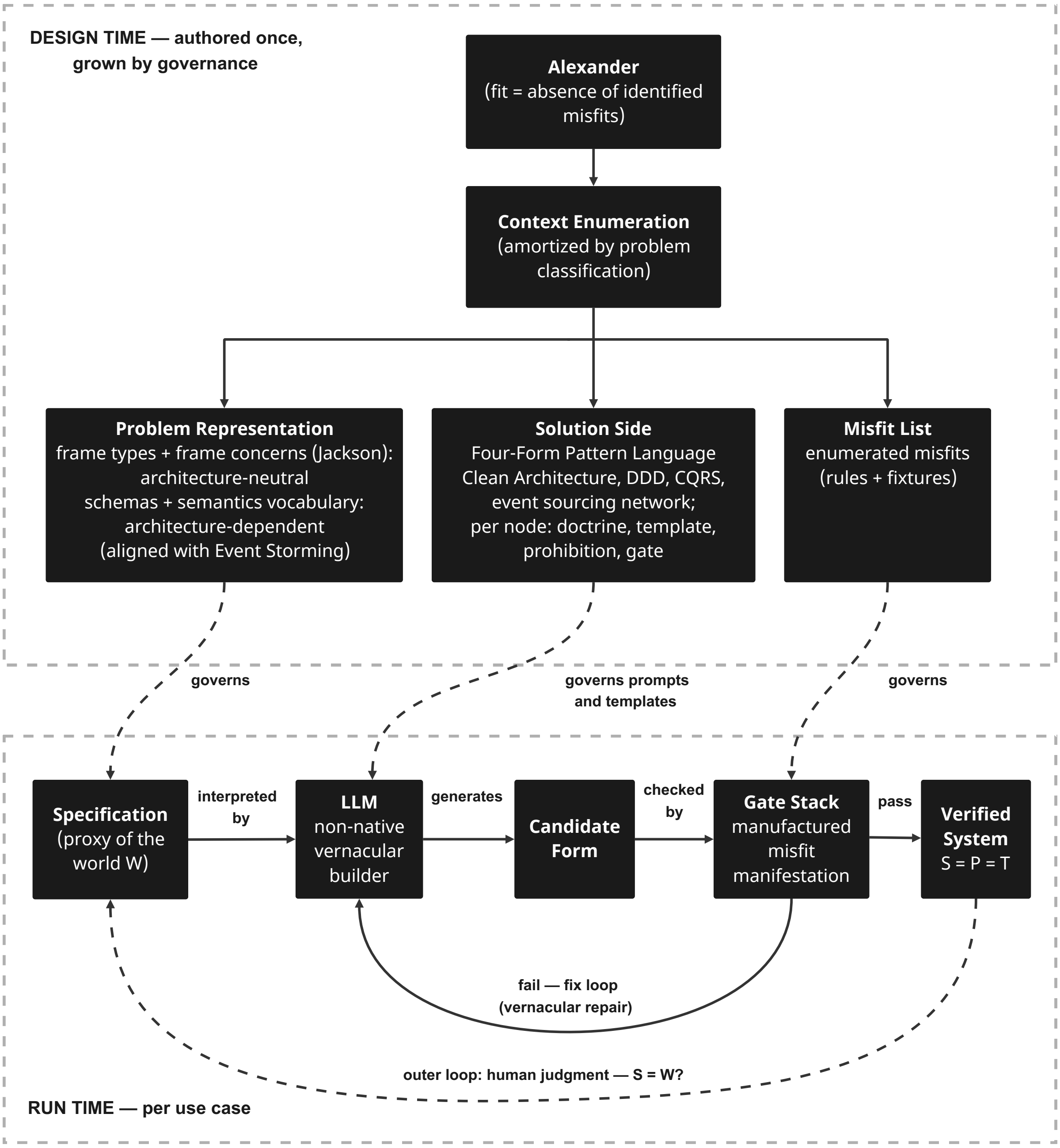

*Figure 1. From theory to machinery. The gate stack and fix loop of the run-time band belong to the harness; downward dashed arrows indicate which design-time artifact governs which run-time stage; the outer loop returns the judgment no gate can render. Together the two bands realize the S = P = T = W assurance model.*

# 3. The Methodology

We name the methodology *Misfit-Governed Development* (MGD): candidate forms are governed by the explicit enumeration, detection, and repair of misfits between the form and its context.

The builder in this paper is a generative actor with a non-native LLM at its core (Section 4.1), and a candidate is admitted only when the harness detects no enumerated misfit.

In this paper, we distinguish the *law* from the *verification harness* (hereafter, the harness). The law consists of the design-time artifacts of Figure 1, namely, the problem representation and the solution-side language, together with the misfit list. Anything that carries normative authority over what the builder produces, but cannot render a verdict, belongs to the law. In practice, the law tells the builder what system to produce and how to produce it in the project's local tradition.

The harness is the run-time machinery that enforces the law, namely, gates, fix loops, reconciliation checks, and their own verification. Anything that can render a verdict or halt progress belongs to the harness. In practice, the harness determines whether the generated artifacts conform to the law. The other run-time machinery an agent needs, such as managing its context or routing its tool calls, is outside our scope. Section 7 relates this focus to the wider harness-engineering literature.

The boundary between the law and the harness is defined by role rather than by execution time. A gate deliberately connects the two sides: its governing rule is authored at design time and belongs to the law — as a pattern's fourth form, or as a specification-side rule (Section 3.4) — while the executable checker that evaluates the rule renders verdicts at run time and belongs to the harness. Every pattern includes its own checker by methodological requirement; rule and checker are authored together so that the executable verdict stays coupled to its governing rule.

At run time, the law is also present, carried into the builder's prompts as guidance. Guidance can inform the builder, but it cannot render a verdict or halt progress. The same distinction applies to the generation workflow of Section 3.3. The prescribed traversal order guides the builder and therefore belongs to the law. The harness does not drive the traversal. Instead, it refuses to conclude one while any gate fails or any step remains unrecorded. It returns each failing verdict to the builder.

Building and maintaining the run-time machinery around an agent has recently come to be known as *harness engineering* [2, 3]. We adopt the term for the verification harness studied here. The novelty is not the term itself but a derived account of what a sufficient verification harness must contain and why (Section 2.5). The governing principle is that verification does not stop at the generated code. The law itself, and the machinery that enforces it, must also be verified — for internal integrity and fidelity to its sources; its fit with the world remains the outer loop's judgment (Section 4.2). This does not regress without end: the chain of checks terminates at a mechanically consultable primary where one exists, or at the human ruling that authorizes the proxy (Section 4.2; Appendix A).

Table 1's derivation can now be distilled into seven governing principles. Where Table 1 preserves the derivation of the obligations, these principles turn their governing consequences into rules for operating the methodology. Each is derived in the sections indicated below, and none is merely advisory. Every one is realized as an enforceable property of the reference implementation, through the gate stack, the structure of the dual-loop harness, or the human-gated legislative circuit (Sections 3.7 and 4.4).

1. **Trust the stack, not the generator.** The generator's fluency has no standing as evidence of correctness (P1; §3.5). A stronger generator improves the odds that a candidate satisfies the harness, but the harness's verdict still determines whether the candidate is accepted. A verdict counts only if the responsible detectors actually ran and their liveness is verified: a skipped gate has not passed (P2; §3.4–3.5, §5). The representations on which that verdict depends are themselves verified and reconciled with their sources (P10; §4.2, §5). And a history of passing verdicts gives no grounds for retiring enforcement: observed compliance cannot establish that the builder has internalized the tradition (P9; §4.5).

2. **Enforcement over guidance.** Prompt-carried guidance cannot be trusted to carry compliance. Only gates are authoritative (P1, P3; §4.1, §5).
3. **In this methodology, a pattern exists only in four forms: doctrine, template, prohibition, and gate.** A pattern lacking any of the four is treated as not existing (P3; §3.4).
4. **A false positive is never fixed in the generated artifacts.** If the gate is wrong, fix the gate. If the gate is right but the law behind it is wrong, revise the law through the legislature of Principle 6. Never modify the generated artifacts simply to make a broken check pass. Otherwise, the error in the gate or in the law behind it propagates into the generated artifacts (P6, P10; §3.7, §4.4).
5. **Detection cost sets the affordable feedback cadence.** Machine-priced detection affords a continuous inner loop. Human-priced detection affords only a budgeted outer loop (P7; §4.3). The separation of cadences is a consequence of this asymmetry rather than a design preference; the dual-loop architecture realizes it.
6. **The builder may not amend the law.** Repairs belong in the generated artifacts, never in the tradition. Amendments pass through the human legislature (P6; §3.7, §4.4). This does not freeze the law. The model may draft and even execute an amendment. It may never approve one.
7. **Machines verify S = P = T; humans judge S = W.** The first three relations are mechanized so that human attention can be devoted to the fourth (P8; §4.2, §6). S = W is the outer loop's irreducibly human relation, not its only responsibility: the same loop also governs the law and the harness on which the machine verdicts depend (Principle 6).

## 3.1 The dual-loop model

Figure 2 presents a dual-loop conceptual model of AI-assisted software development. A human designs the law, a representation of both the problem and the local tradition, and iteratively amends it as new misfits emerge during verification and use. An LLM performs a *non-deterministic transformation* of this representation into candidate code. The candidate then enters an *inner loop*. The loop runs inside an autonomous agent zone, iterating the candidate against deterministic verification by tests, contracts, and static checks. A human may help (say, approving an operation or answering a question about the specification) but never judges the candidate. Within this loop the specification is *provisionally assumed correct*. A failed verdict provisionally blames the candidate, never the specification, and triggers automatic retry.

The verified solution is executed by a machine (the runtime and executable contracts), producing observable behavior. An *outer loop* returns observations of that behavior to the human, who revises the governing representations. At a slower cadence, the same loop also returns observations of the system's lived consequences. When failure persists across repeated inner-loop retries, the loop stops blaming successive candidates and escalates to the outer loop for diagnosis.

The revised artifact is not always a specification. Sometimes the tradition itself must be revised. An escape, a defect that slipped past every gate, triggers diagnosis of the governing chain: if the law already represented the violated expectation, the escape reveals a detection fault — a detector missing, ineffective, unreachable, or reading a stale proxy; if it did not, it reveals an omission in the law or a demand newly discovered in the world (Section 4.4). A false positive, a gate flagging code that should not have been flagged, reveals a fault in the gate when the flagged code conforms to the law; when the gate is right but the law behind it is wrong, the fault lies in the law (Principle 4). In both cases the remedy lies outside the generated artifacts and flows through the legislative circuit of Section 3.7. Either way, revision happens at design time, never inside the inner loop. Every verdict is rendered against expectations, and the expectations live in the governing representations, specifically in the specifications and the misfit list. Because humans author these representations, verification can be performed deterministically by machines while accountability remains with humans.

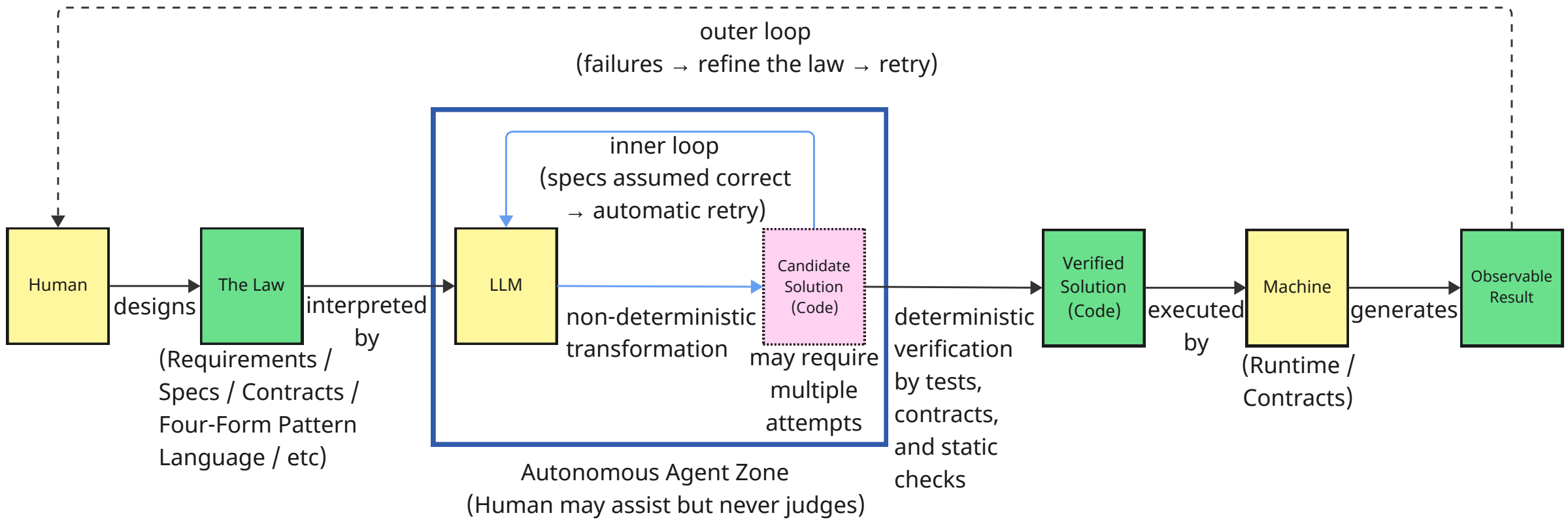

*Figure 2. A dual-loop conceptual model of AI-assisted software development. The inner loop drives the machine-decidable relations (S = P = T); the outer loop carries the one judgment the machinery cannot make on its own authority (S = W) and governs the law and the harness.*

The two loops differ in their observers and in their tempo. The inner loop observes machine-checkable verdicts and runs at machine cadence. The outer loop observes the system's behavior in the world and runs at whatever cadence human attention can be budgeted for. Section 4.3 derives this division from first principles.

## 3.2 The problem representation: problem-frame specifications

The problem side is written in a structured specification language organized by frame type. Each use case, once classified under a frame type, is specified by five kinds of artifact, stored in its own directory at `{project}/{bounded-context}/aggregates/{aggregate}/{frame-type}/{use-case}/`:

1. `frame.yaml`, which names the frame type, the problem-world facts, and the frame concerns.
2. Requirement files, which carry the description, the rationale, and the acceptance criteria that verify them.
3. A machine specification, which defines the machine's responsibilities, the use case with its input and output, and the controller that exposes it as an endpoint.
4. A domain specification, which is frame-type-specific. For state-mutating frames it carries the aggregate part with its attributes, domain events, invariants, and Design-by-Contract clauses [13]. For the others it carries the read model, the transformation's input/output contract, or the trigger wiring.
5. `acceptance.yaml`, which states the scenarios in Given–When–Then form, with full traceability to the requirements.

These artifacts are written per use case, not all in advance. A use case can be analyzed, specified, and generated on its own, and its domain specification carries only the part of the aggregate that the use case touches, which is why that artifact is named the aggregate *part*. The specification therefore grows piece by piece with the system, in Alexander's manner of piecemeal growth [9], rather than being completed up front.

Seven frame types are currently distinguished. Five are inherited from Jackson's catalog [12]: commanded behavior (CBF), simple workpieces (SWF), required behavior (RBF), information display (IDF), and transformation (TF). The sixth, reactive invocation (RIF), is an addition of ours in which a triggering event or schedule invokes a further operation, in this architecture most commonly to maintain eventual consistency between aggregates (Section 3.3). The seventh, the connection frame (CF), is an extension of ours that elevates the role played by connection domains in Jackson's framework [12] into a frame type of its own. It specifies the connection between separated processes, which this architecture realizes as infrastructure relays carrying events between them, and it lies outside the application tree and therefore outside the empirical counts of Section 5.

The specification language itself has machine-checked syntax and partially machine-checked semantics. JSON Schemas constrain the core file types. A 39-rule specification linter enforces format, cross-file traceability, structural well-formedness, and cross-frame value-object consistency. A controlled *semantics vocabulary* of declarative tags (e.g., `value_immutable`, `identity`, `soft_delete_flag`) maps each tag to mandatory code-generation consequences (no setter generated, invariant clause injected). The problem representation itself, namely, the frame catalog, the schemas, the linter rules, and the semantics vocabulary, is amendable only through the human-gated legislative circuit of Section 3.7.

The tree above the frames is organized by domain-driven design's decomposition units (bounded contexts and, within them, aggregates [14]). As a result, the problem-side decomposition and the solution-side packaging use the same units. An aggregate is therefore both a unit of specification and a unit of generated code. This organization belongs to the representation, not to the frame types. The problem representation is architecture-dependent. The frame types themselves classify the problem and are architecture-neutral. The representation built on them expresses the stakeholder demands in the vocabulary that the chosen solution family requires.

Before the specifications are written, an Event Storming workshop [15] brings domain experts and developers together to model the business flow as a timeline of domain events. From that timeline, commands, policies, read models, aggregates, and bounded contexts are identified, and a class diagram of the domain is drawn. The frame types are deliberately aligned with Event Storming's grammar; the mapping that follows is our classification scheme, not part of Jackson's catalog. Task-based commands, those that ask the system to carry out a business activity, become commanded behavior frames. Commands that simply establish a state become simple workpieces frames. Read models become information display frames. Policies ("whenever this event occurs, do that") are classified by their reaction. A policy that invokes a command becomes a reactive invocation frame; one that maintains a read model becomes a transformation frame. A rule that obliges the system to act on its own, when a condition holds rather than when a user commands, becomes a required behavior frame. The aggregates that organize the specification tree are discovered on the Event Storming wall.

Once a unit is classified under a frame type, it inherits that type's frame concerns, namely, its pre-enumerated misfit list (Section 2.4). Event Storming is thus the methodology's structured first contact with the primary problem context (Section 4.2). The problem-side misfits that the frames later formalize have two sources: the generic concerns come with the frame type, and the domain-specific demands that instantiate them originate with the domain experts, although the specification process may reveal omissions that require further elicitation. Figure 3 shows a classified slice of such a wall.

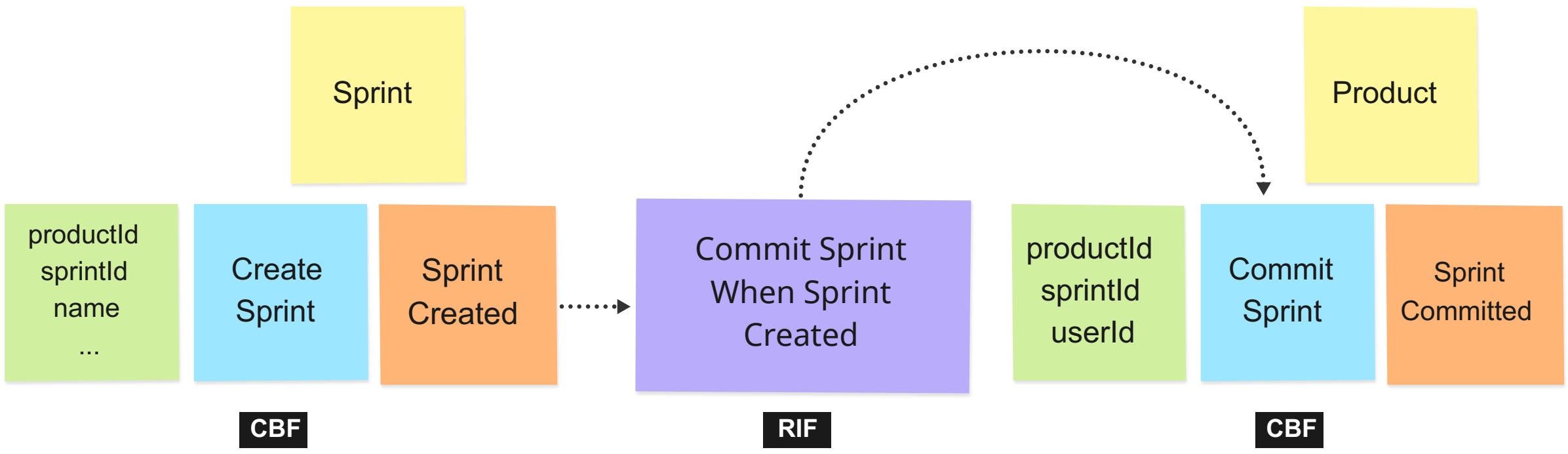

*Figure 3. From the Event Storming wall to problem frames: a slice of the board, classified. A command with its input and event becomes a commanded behavior frame on the Sprint aggregate; the policy "Commit Sprint When Sprint Created" becomes a reactive invocation frame wiring that event to a command on the Product aggregate — a second commanded behavior frame. The dotted couplings denote the eventual-consistency links between aggregates (Section 3.3).*

## 3.3 The solution-side language: a pattern network

The solution-side language is a network of patterns that holds the chosen architecture fixed. It was *drawn* before any prompt, gate, or workflow existed. The network does not prescribe every form the builder may produce. Figures 4–8 show only the architecturally significant part of the network, not the complete language. Design- and implementation-level patterns are introduced later, in the templates and rules attached to each node of the pattern network (Section 3.4). Those templates and rules incorporate cataloged design patterns (e.g., [14, 16, 17, 18, 19]) together with the team's local coding conventions.

The network itself is an engineering choice in the sense of Section 2.5, not part of the theory. Alexander expected this diversity, for every living culture has its own pattern language [11]. This network is our team's architectural culture made explicit. The culture itself is not new. It has accumulated since 2018 through the continuous development of ezKanban, a Kanban system we built with domain-driven design, Clean Architecture, event sourcing, Command Query Responsibility Segregation (CQRS), and design by contract [20].

The network is also selective. It includes a pattern only when the team constrains how it is used. The builder knows many more from training. If the team does not constrain a pattern, the network simply leaves it out. The builder stays free to use it, within the fixed architecture. This does not freeze the architecture. The team may still change the network through the legislative circuit of Section 3.7, and because every pattern exists in four coupled forms, changing a pattern also means changing its gate (Section 3.4). The derivation obliges only that some such language exist and be enforced (Table 1, P3); which architectural family it encodes is revisited in Section 8.

This section presents the macro-structure and its rules, and then unfolds each layer into its own pattern map. Finally, it explains the locality this structure supplies and the order in which workflows traverse it.

**Macro-structure and rules.** The macro-structure is Clean Architecture [21], whose principles we operationalize in the pattern language as three named rules (Figure 4):

1. the *Layering Rule*: every artifact lives in one of four layers, Entities, Use Cases, Interface Adapters, and an IO layer (our name for Clean Architecture's Frameworks and Drivers).

2. the *Dependency Rule*: source dependencies point inward, from the IO layer toward the Entities layer.
3. the *Cross-Boundary Rule*: layer crossings pass through ports, and domain objects themselves never cross. Inside the Use Cases layer, they are converted into a data transfer object toward the adapters or a persistent object toward storage. Incoming data structures are converted back into domain objects.

In the implementation, the Spring Boot entry class belongs to the IO layer's Main component. It sits in the project's root package so that Spring Boot's default component scan includes all packages below it. All three rules are machine-checked (Section 3.5). The layering is enforced by the structural gates, which pin each generated artifact's package. The dependency and cross-boundary rules are checked by architecture tests.

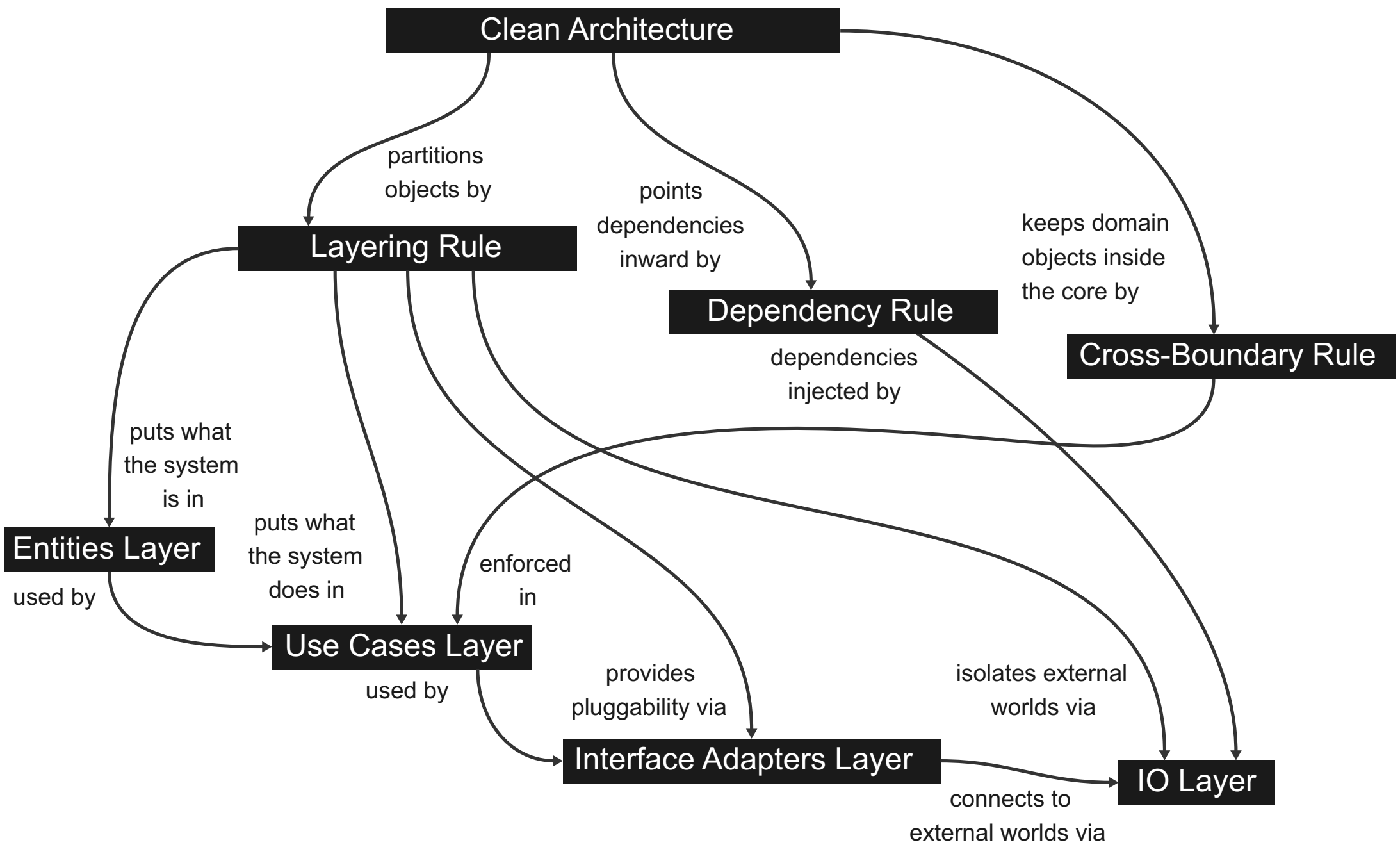

*Figure 4. The macro-map of the solution-side pattern language: Clean Architecture realized as three named rules, and the four layers they govern.*

**The layer maps.** Each layer then unfolds into its own pattern map, in the manner of *A Pattern Language*'s ordered networks. In such a network, every pattern completes the larger pattern above it and is completed by the smaller ones below it [11]. A gray node marks a pattern that lives in another layer's map. Through these shared nodes, the four maps interlock into one network.

The Entities-layer map (Figure 5) defines the event-sourced domain vocabulary, built on domain-driven design's tactical patterns [14]. Aggregates are composed of Entities and Value Objects. A Read-Only Entity lets an aggregate hand its internal entities to outside clients in a form that answers queries but rejects modification, preserving the aggregate's encapsulation [18]. Domain Events are the recorded state changes from which an aggregate's state is rebuilt by replay. At the bounded-context boundary, Domain Events split into Internal Domain Events and External Domain Events [18].

The Use-Cases-layer map (Figure 6) encodes CQRS [19] as grammar rather than advice. Command, Query, Reactor, Projector, and Notifier use cases are distinct pattern families. On the command side, a Repository [14, 17, 19], implemented as a bridge [16], converts the Aggregate to persistent objects *inside* the Use Cases layer, so the Aggregate never crosses outward. The Repository supports both persistent forms, current state and a stream of domain events. Three further persistence patterns are deliberately kept distinct. Projection serves the Query use case from the read model. Inquiry supplies command-side use cases with facts needed for decision making beyond the Repository's lookup of the aggregate being acted upon, including facts derived from other aggregates. Archive is the port through which projectors (themselves instances of the transformation frame) store the read models. The remaining port, gateway, decorator, and boundary-helper families are enumerated in the figure. Typical members include idempotency, validation, authorization, and transactional decorators on the in-port side; circuit-breaker, caching, and logging/tracing decorators on the out-port side; TCP sockets, SSE/WebSocket, and message brokers as message push gateways; e-mail, payment, and third-party APIs as business capability gateways; and time, ID generation, and randomness as ambient capability gateways. Repository Adapters include, for example, the outbox repository adapter and the event-sourced repository adapter.

The Interface-Adapters-layer map (Figure 7) fills both sides of the Use Cases layer's ports. The inbound adapter families admit the outside world's inputs. REST controllers carry commands and queries, the scheduler adapter fires scheduled reactions, and message consumers deliver the events that trigger reactors. The outbound adapter families mirror every out-port declared in the Use Cases layer. Each adapts an infrastructure component in the IO layer. The persistence adapters realize the Repository, Projection, Inquiry, and Archive ports. The capability and message-push adapters serve the remaining out-ports.

The IO-layer map (Figure 8) completes the unfolding. The Main component, composed of Bootstrap and the Composition Root, injects the other layers' dependencies. The infrastructure components, such as the embedded HTTP server, database drivers, and message broker clients, connect to systems outside the architecture.

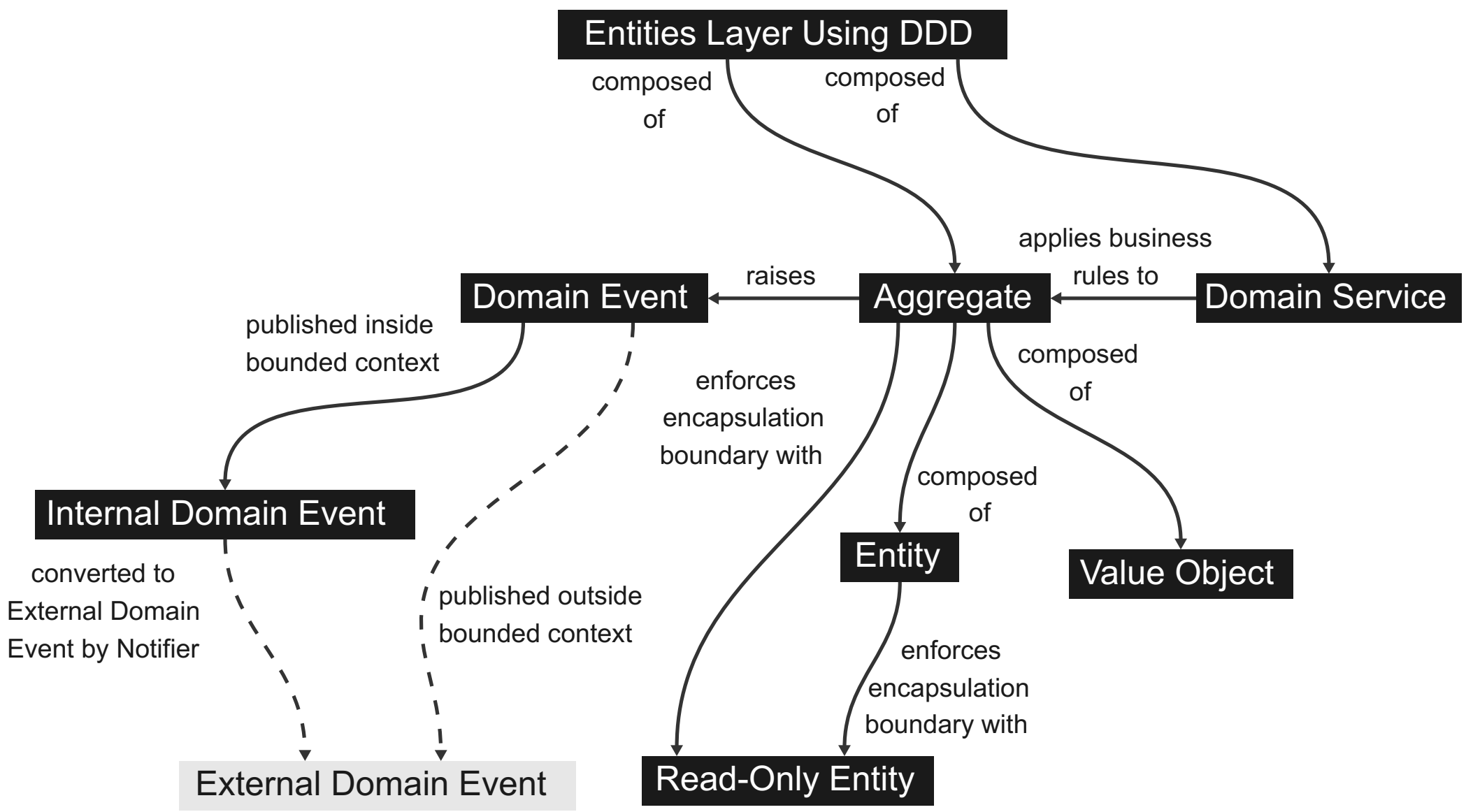

*Figure 5. The Entities-layer pattern map: the event-sourced domain vocabulary. The gray node marks a pattern that lives in another layer; here, External Domain Event belongs to the Use Cases layer, where a Notifier converts internal events into external ones.*

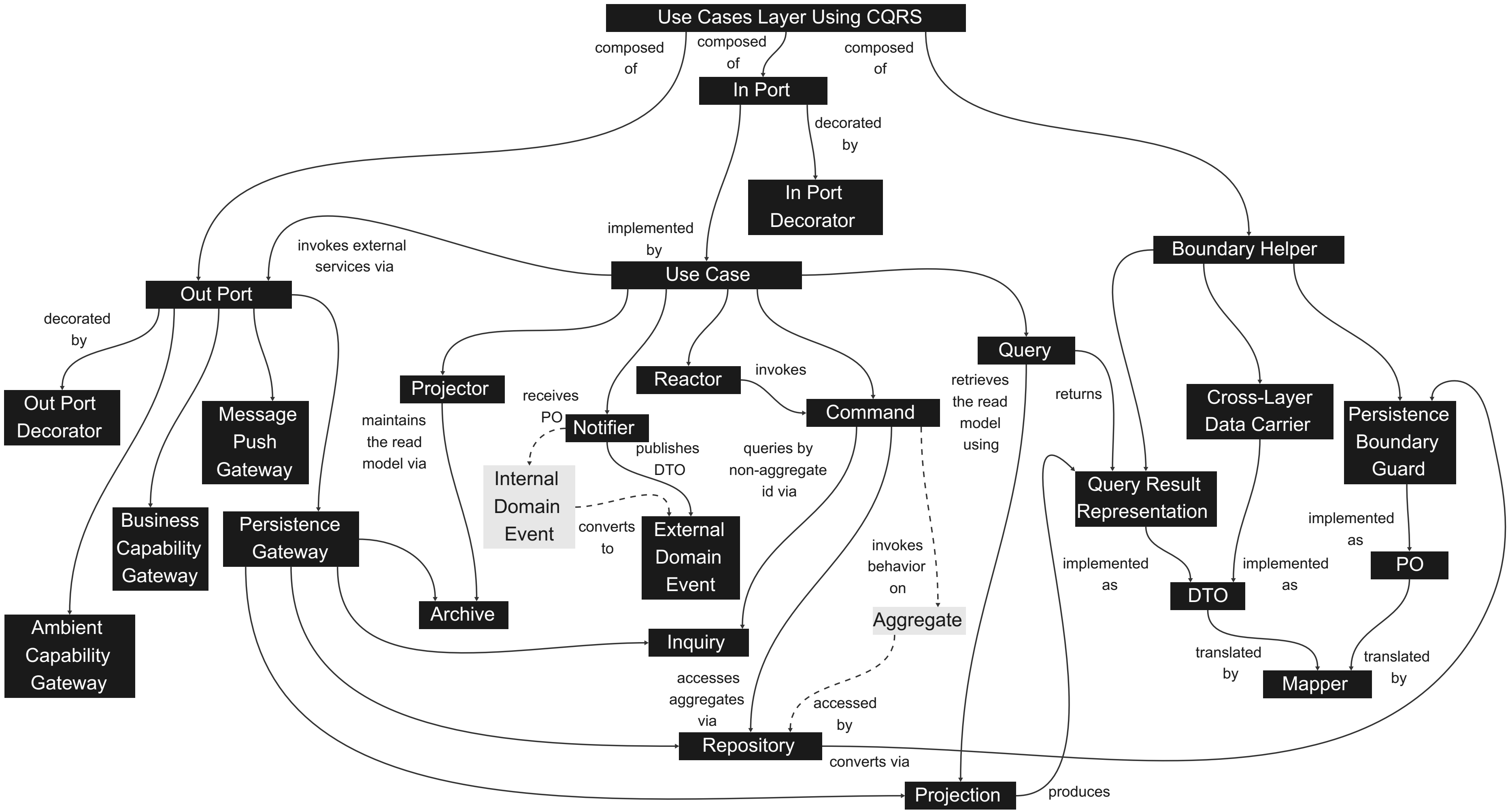

*Figure 6. The Use-Cases-layer pattern map: CQRS encoded as grammar — command/query/reactor/projector/notifier families behind in-ports, a taxonomy of out-ports, the bridge-pattern Repository that keeps the Aggregate inside the layer, and cross-cutting decorators. Gray nodes are patterns from other layers; here, Internal Domain Event and Aggregate live in the Entities layer. Unlabeled arrows read "specialized as."*

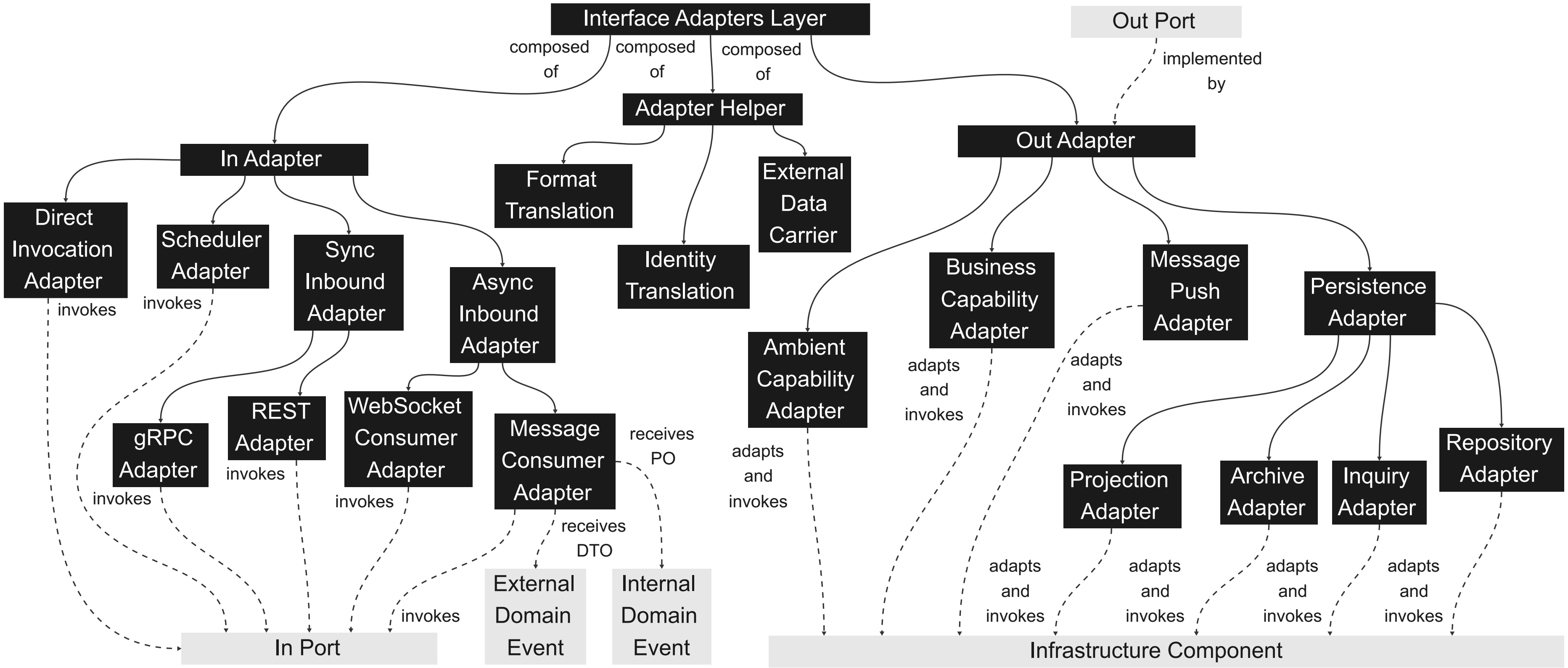

*Figure 7. The Interface-Adapters-layer pattern map: inbound adapter families invoking the Use Cases layer through its in-port, and every out-port mirrored by an adapter family that adapts an infrastructure component in the IO layer. Gray nodes are patterns from other layers. Unlabeled arrows read "specialized as."*

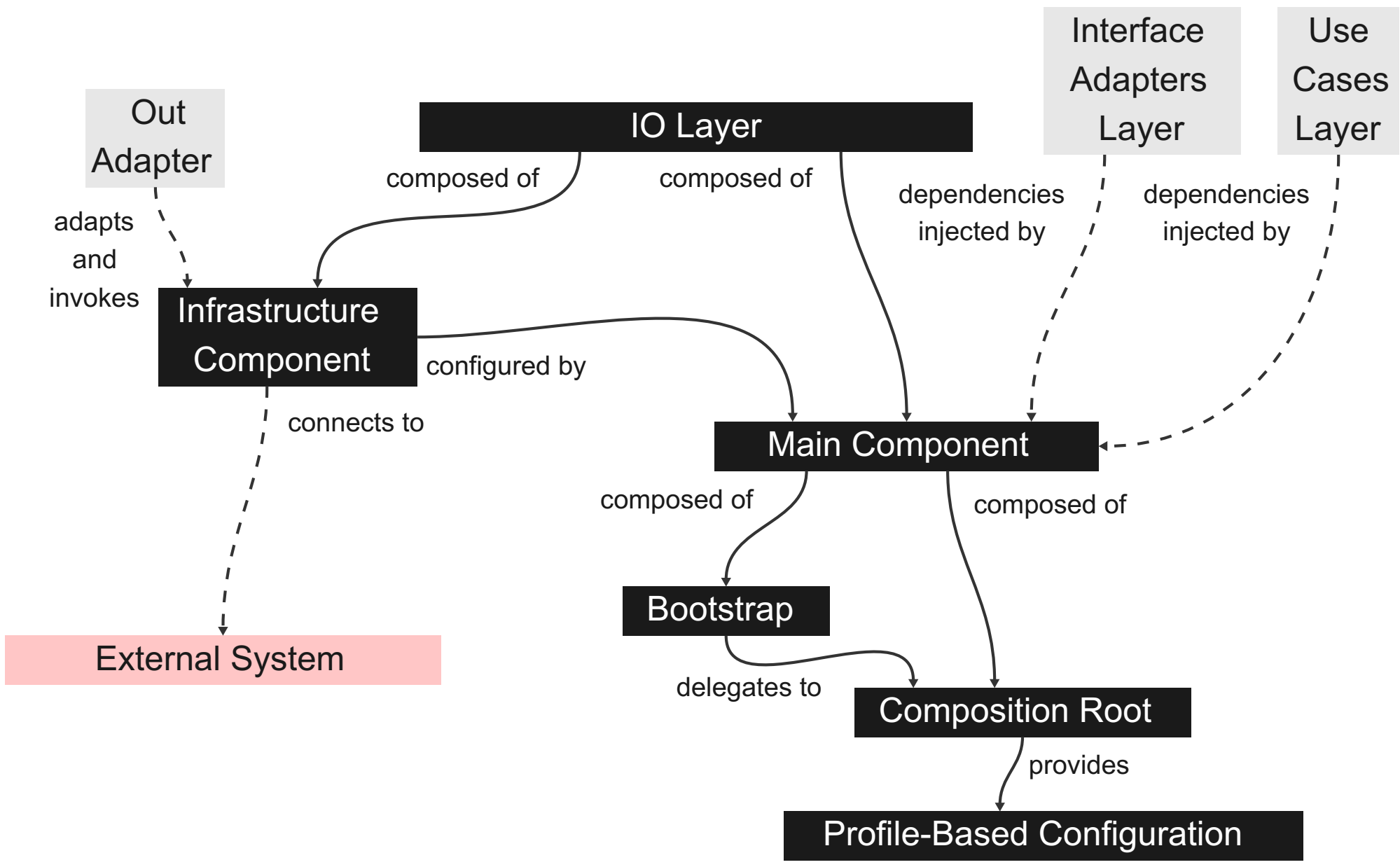

*Figure 8. The IO-layer pattern map: the Main component, composed of Bootstrap and the Composition Root, injects the other layers' dependencies and configures the infrastructure components, which connect to external systems. Its Composition Root provides the profile-based configuration behind the two infrastructure profiles (Section 3.5). Gray nodes are patterns from other layers; the pink node is an external system outside the architecture, such as an e-mail server, a database server, a message broker, or a payment gateway.*

**Locality.** Beyond vocabulary, these boundaries supply the *locality* that a homeostatic process requires. Bounded contexts partition the misfit network into weakly coupled subsystems, and aggregates draw the boundaries within which repair is intended to be local. Domain events are the deliberately weak coupling between subsystems. Consistency is transactional inside an aggregate and *eventual* between aggregates, restored by reactive invocation frames that consume one aggregate's events and invoke operations on another [19].

Clean Architecture's structural rules add the vertical counterpart to this horizontal decomposition. Dependencies point only inward, and layers meet only at ports. These boundaries constrain how far a repair's effects can propagate, although they cannot force the builder to keep its edits inside them; the regression guard of Section 3.6 covers that residual risk. Neighboring layers touch the repaired aggregate only through its ports, and neighboring aggregates hold only its identifiers. In practice, this lets different aggregates be generated and repaired in parallel, while the methodology serializes work within one aggregate to preserve a stable repair context. The identifier value objects that cross aggregate boundaries are generated from a standard form and skipped if they already exist, so parallel runs do not collide on them. Locality is necessary but not sufficient for homeostasis (Section 4.3 completes the argument).

**The authoring sequence and its traversal.** The pattern language was drawn first as a network. Each node in use then received its doctrine, template, prohibitions, and gate (Section 3.4's four coupled forms are per-node obligations). Finally, each problem-frame type received a workflow that is an *ordered traversal of a subgraph of this network*. The traversal sets the order in which nodes are generated, not the runtime call flow. For example, a commanded behavior frame unfolds Entity-first: the Aggregate with its Domain Events, then the Command use case with its ports, then the repository and outbox adapters, and then the controller. An information display frame unfolds from the read model through its Projection to the Query use case and its controller. A reactive invocation frame wires an existing trigger event to a generated Reactor and its target operation. Every traversal ends in the IO layer, where a configuration class in the aggregate's own package wires the new parts into the running system (Figure 8). The composition root discovers that class through component scan, so parallel traversals never touch a shared file.

Drawing the language before writing any prompt, and enforcing it through each pattern's gate, gives the generated system its formal morphological coherence; Section 6 separates this formal sense from Alexander's felt wholeness. Individual prompts never negotiate their mutual fit at generation time, because their intended compatibility was specified once, at the level of the language.

## 3.4 The four-form pattern language

The language's reader is an unreliable generator rather than a judicious professional. To govern that reader, the methodology requires each pattern in the network to be realized in all four coupled forms before the builder may generate from it. Together, these forms define what we call a *four-form pattern language*:

1. **Doctrine**: prose stating when the pattern applies and why.
2. **Template**: the positive form, namely, code skeletons precise to the import path. The builder fills in the domain content. Everything else is supplied by the template.
3. **Prohibition**: the negative constraints, namely, what must not appear and what must not be absent (95 always-active rules at the time of writing). A prohibition typically bans the mainstream idiom that the pattern replaces. A required construct is expressed as a prohibition on its absence.
4. **Gate**: a deterministic checker, with a suite of good/bad fixtures showing that the checker still detects what it claims to detect, and real invocations showing that it still reaches what it claims to scan.

The four forms divide the labor rather than duplicate it. Doctrine explains when and why a pattern applies. The template provides its positive form. The prohibitions encode the local departures from the builder's learned defaults and serve as the repair anchor for reported violations. The gate alone renders the authoritative verdict. Like every mechanism in Table 1, the four-form language is one way of discharging its obligation (P3), not the only way (Section 2.5).

A pattern that exists only as prose is treated as not existing, because prose guidance alone does not reliably constrain the builder (Section 5); several drawn nodes of Section 3.3 remain reserved vocabulary for future growth, not yet existing in this sense.

Not every gate is the fourth form of a pattern. The pattern network is the solution-side representation (Section 3.3), and the four forms govern it. The gate stack also enforces the problem side (Section 3.2). A rule may require that the receiving service read every input field the specification declares, so that no input is silently dropped. Such a rule asks whether the code discharges an obligation the specification states, rather than whether a construct has the right shape, and the authority it cites is the specification schema rather than any pattern.

We keep these rules outside the pattern language rather than folding them into it. Both the pattern network and the specifications are proxy contexts in the sense of Section 4.2, but they proxy different things. The pattern network carries how the local tradition builds. The specifications carry what this particular system must do. A gate that checks the second belongs to the S = P or S = T relation of the assurance model (Section 4.2; Appendix A.2), not to the language. The harness runs both kinds in one stack, but that is an implementation convenience. Treating every gate as a pattern's fourth form would blur what a pattern is.

The language also keeps its own rules consistent. The same rule often appears in several places, including a pattern's doctrine, the guidance the builder reads, and the gate's rule catalog. Different consumers require different representations, so the language restates each rule where it is consumed.

To prevent these copies from drifting apart, each rule has one designated original, recorded in a single-source-of-authority registry. The other occurrences restate the rule in full rather than merely pointing to it, because the builder must see the rule during generation; machine-readable links tie each restatement to its original, and consistency checks fail until a diverged copy follows its original. At the start of every session, before the builder generates anything, self-consistency checks recompute facts asserted in more than one place and fail when two sites disagree. Retired doctrine is governed the same way: superseded wording can reappear as guidance, so a *doctrine anti-pattern registry* records it and a scanner flags it wherever it resurfaces, in effect automating the forgetting of retired doctrine. Deletion happens once; forgetting must be enforced continuously. All of this maintenance belongs to the outer loop (Figure 2): the checks detect mechanically, but every amendment to the language is approved by a human.

## 3.5 The gate stack: manufactured manifestation

At the time of writing, generation passes through 28 blocking gates of five kinds:

1. a specification-quality gate: the linter of Section 3.2.
2. a dual-profile test gate: every generated test suite, together with the permanent architecture tests of Section 3.3, must pass under both an in-memory and a full outbox/PostgreSQL infrastructure profile.
3. eighteen per-use-case structural gates: for each use case, they check that the generated code and tests stay structurally faithful to its specification.
4. seven aggregate-wide closeout gates covering spec-boundary, deliverable-manifest, encapsulation, and cross-frame consistency checks. Cross-aggregate checks do not form a kind of their own. Distributed across the dual-profile, structural, and closeout kinds, they cover globally unique endpoints, inter-aggregate import isolation, and the correct use of value objects and identifiers that belong to another aggregate.
5. a deterministic-review gate applying 188 machine-checkable rules. For example, one rule forbids `Instant.now()` in domain code, so that time always comes from an injected provider and tests stay deterministic.

Two further gates sit outside the blocking count: advisory gates whose warnings (over-specification and unnormalized optional inputs) surface but never block. One further mechanism is not a gate at all: an optional multi-model semantic review, in which three LLMs from different vendors review the generated code against the specification and an arbiter LLM then resolves their disputed findings.

The gates compare the code against two different sets of expectations. The structural gates compare it with the problem-frame specifications for each use case, while the closeout gates compare it with the same specifications across the whole aggregate. The 188 deterministic-review rules mostly compare it with the solution side: framework contracts, architectural doctrine, and platform quirks, which are the same for every use case. The split mirrors the two languages of Sections 3.2 and 3.3 (Section 4.2 develops the point).

The eighteen structural gates fall into three broad families. Each family reads a different part of the specification and demands the matching elements in the generated code and tests. *Coverage gates* check that every specified scenario not deferred by the specification, and every contract clause, has a corresponding test, that no test cites a scenario the specification does not declare, and that each test asserts the kind of outcome its scenario declares and mentions the states and quoted values the scenario names. *Field gates* verify *field-level agreement* across representations. A field declared in the specification must reappear, correctly named and typed, throughout its journey. Depending on the field, the journey spans value objects, entities, events, mappers, persistent objects, and data transfer objects (DTOs). The field must be genuinely carried at every hop. The converse check, that no field is realized without being declared, is blocking for domain-event records, whose fields enter the permanent event log, and advisory elsewhere. A companion rule in the deterministic-review gate blocks the same kind of addition on the aggregate root: a public behavior method, or an event such a method applies directly, that the specification does not declare. *Controller gates* extend the same discipline to the HTTP boundary. They check request DTOs, endpoint semantics, error handling, and controller-test completeness. A field that vanishes mid-chain (an input silently dropped by a service, a derived value lost in reconstruction) is invisible to any single-layer check, because every layer remains locally well-formed. It is detected only by checking the chain hop by hop, field by field.

Every gate is deterministic. The determinism is essential, not incidental. A gate is an ordinary program, a parser, a pattern matcher, or a test run. No language model sits in any gate's verdict path. The inner loop has no human in it to arbitrate, so a gate must return the same verdict on the same artifact every time, given the same pinned inputs and environment (Section 3.6). A detector that sometimes flags an artifact and sometimes passes it forces the judgment call that the zone was designed to avoid (P8). A repair loop cannot reliably converge when the same artifact receives different verdicts. Each fix round responds to a different verdict, and the violation count can no longer show whether the repair is making progress. The vernacular's physics was deterministic in this sense: the roof leaks every time it rains.

The optional multi-model review is non-deterministic by design. The harness's deterministic predicates cannot decide whether the generated code faithfully reflects the specification's *intent* beyond what they already check (Appendix A). The review therefore estimates this *semantic faithfulness* from the assessments of three LLMs from different vendors rather than computing it directly. The review remained advisory throughout the study and accumulated little operational history, so this paper reports it as part of the design, not as evaluated evidence.

## 3.6 The fix loop: mechanized vernacular repair

Gate failures do not end the process but enter a three-layer fix loop. The layers handle test failures (the dual-profile suite), structural violations (batched across the structural gates), and pattern violations (the deterministic-review rules). A repair round batches violations within one layer only, and each layer has its own repair instructions. The optional multi-model review stays outside the loop because a non-deterministic verdict cannot anchor convergent repair (Section 3.5). Its findings go to human adjudication instead. Five properties distinguish this fix loop from a naive retry loop:

1. *Batching*: re-verification is cheap because it only runs ordinary programs, while repair is expensive because it asks the builder to rewrite the code. Each round therefore collects every violation the layer reports and fixes them together. A dozen violations cost one repair round rather than a dozen, and violations that share a root cause are fixed as one. The vernacular process offered no comparable economy. Misfits revealed themselves through continued use of the building and were repaired as they appeared.

2. *Tradition-anchored repair*: violations reach the builder not as bare verdicts but with the offending code, the violated rule, and a pointer into the tradition. A gate failure thus identifies the violated pattern and re-exposes its prescribed form as the repair target. Before fixing, the builder is instructed to re-read the authoritative pattern document for each rule it violated. Repair is thus guided by the tradition rather than by the model's learned defaults. This matters because the risk of drifting back to mainstream code is highest during repair. By the time repair begins, the patterns read at generation time may no longer be present in the context window, and an unanchored model falls back on mainstream habit. The re-read instruction is guidance, not enforcement. A repair that ignores it and drifts back to the mainstream simply fails the gates again on the next round.
3. *Regression guarding*: every repair round re-runs prior gates, because an unreliable builder's fixes are not guaranteed to be local.
4. *Stall detection*: if the violation count fails to decrease for three consecutive rounds, the loop halts and escalates to the human. This is an explicit valve. A problem that mechanical repair cannot resolve is handed upward as a design question. In Alexander's terms, mechanical repair below the valve plays the unselfconscious role, and design judgment above it is self-conscious (Sections 2.2 and 2.3). The same escalation can happen before generation begins. If the builder cannot interpret the specification, the process halts and raises the question to the human rather than answering it by guesswork. The vernacular process had no corresponding valve. A repair that failed remained a problem for the culture itself rather than being escalated to a separate design authority.
5. *Context freezing*: the context against which the generated artifact is judged remains fixed throughout a fix-loop window. Inputs under the workflow's control, such as the specification and the framework version, are pinned when the loop begins. The tradition and the verification machinery cannot change from inside the loop and evolve only through the separate human-gated legislative circuit (Sections 3.7 and 4.4). This stable context allows Alexander's one-at-a-time process of correction to converge (P4). Fixes can land one at a time only if the conditions against which they are judged are not changing beneath them. Stability makes convergence possible, not guaranteed; the stall valve of property 4 covers the difference.

Together, the five properties mean that no single repair round has to fix everything correctly. A round needs only to reduce the violation count and to avoid breaking what already passed. The gates check both conditions cheaply. The loop therefore tolerates imperfect rounds but escalates when progress stalls.

## 3.7 What the builder may not touch

In every vernacular culture, repair experience flowed back into the tradition. What the craftsman learned fixing the roof passed slowly, over generations, into that tradition. This channel is deliberately severed. The LLM fixes the generated artifact at hand. It is never permitted to modify templates, rules, or gates on its own authority.

The prohibition is enforced rather than requested (Principle 2, enforcement over guidance). The builder cannot write to gate verdicts or workflow state. A guard rejects any attempt to do so. Otherwise, the builder could simply rewrite the evidence that says whether its work passed. Tradition files are protected differently. The builder can physically edit them, but such an edit has no legislative standing: consistency checks detect the change and flag any dependent document that no longer agrees. The same checks run again at the start of every session, together with the doctrine-registry checks. An unauthorized change therefore cannot become part of the tradition on the builder's authority alone: divergent copies are flagged, and even a consistent rewrite acquires no legislative standing until a human accepts it. If this boundary depended only on telling the builder not to cross it, it would be guidance, not enforcement.

The tradition can still change, but only through the separate, human-gated legislative circuit. Changes come from three sources: defect post-mortems that must examine what the gates missed (Section 4.4), improvements noticed during generation and saved for later batch review, and audits of the verification artifacts themselves (Section 5). The model may make the edits, but a human decides whether to accept them.

The same circuit handles errors in the gates themselves. If a gate rejects code that actually conforms, the gate is fixed. The code is not changed just to make a broken gate pass. Section 4.4 explains why keeping this boundary is necessary, not just cautious.

# 4. Theoretical Analysis

This section develops the theory as a chain, each subsection answering the question the previous one leaves open: what the builder is (4.1), what it is measured against (4.2), how often the measurement can be afforded (4.3), who may amend the standard (4.4), and whether the scaffold can ever come down (4.5).

## 4.1 The non-native vernacular builder

The vernacular builder's reliability came from enculturation in a single internalized tradition. Individual variations could be poor, but visible misfits were corrected before they became established in the tradition. In an AI coding system, the builder is the generative actor that produces and repairs the software, not the language model in isolation; in our implementation, it comprises the LLM together with the agentic machinery through which the model acts. The LLM at the builder's core lacks these conditions. It is trained across many traditions, so its learned defaults reflect many coding idioms rather than any one local dialect. The builder therefore begins with no special commitment to the local tradition. We model these deviations not as random noise but as a persistent pull toward mainstream patterns learned during training. We call this pull the *bias current*. Because in-context instruction does not change the model's learned parameters, the local tradition is supplied as context rather than internalized. Context can shift behavior, but it cannot show that the learned defaults have been replaced, so the methodology treats the pull as still present. We name this model of the generator the *non-native vernacular builder*, a craftsman fluent in many villages' building styles but native to none, playing the vernacular builder's role without the vernacular builder's enculturation. The harness governs what this actor produces, and its verdicts do not depend on which internal component caused a deviation. The prohibition catalog records recurring deviations consistent with the bias current. Many of the forbidden constructs (`@Service` annotations, `@ActiveProfiles`, `Instant.now()`, ad-hoc repository methods, `javax.*` imports) are common Spring Boot or Java idioms. The catalog marks where mainstream idiom conflicts with local doctrine.

Three design consequences follow:

1. *Enforcement must dominate guidance*: the local tradition supplied in the prompt competes with the model's learned defaults. As the context grows or is compacted, that guidance cannot be trusted to remain effective. Prose therefore guides generation, but gates decide whether the result is acceptable.
2. *Rules guarding anti-mainstream doctrine never retire on statistical grounds*: a quiet period does not mean that the underlying tendency has disappeared. Observed compliance does not establish that the model has internalized the local rule, and a new model may introduce a different set of recurring violations. Every rule must therefore remain demonstrably alive *and reachable*. It must still fire on a bad fixture, and it must still be evaluated on the artifacts a real invocation scans. A rule retires only when a change in the local tradition or architecture removes the situation it checks.

3. *The bias current and the capability share a source*: the training that creates the pull toward the mainstream also supplies the builder's general competence in Java, Spring Boot, and domain-driven design. The local tradition therefore need not restate shared knowledge; it encodes the local choices and constraints that the builder's learned defaults do not reliably supply. This lets a tradition of tractable size govern generation.

The generator is globally fluent but locally non-native. The remainder of the methodology is an engineering response to this asymmetry.

## 4.2 Primary and proxy context

The builder model says what the generator is. The next question is what it is measured against, and therefore where the authority for that measurement comes from. Alexander's context is "that part of the world which puts demands on this form" [4]. Designers, however, rarely work from that world directly. They work from mental pictures of it, and those pictures can distort the structure of the problem. *Notes* therefore distinguishes the actual world from the designer's mental picture and, in turn, from a formal picture in which the problem is expressed as explicit misfit variables [4]. Our problem-frame specifications, taken together, play a similar representational role.

The methodology requires one further distinction. Alexander distinguished the world from its representations. The world ultimately determines whether the system fits, but machines can check the system only against representations of that world. A representation may be internally consistent and still fail to represent the world correctly. We therefore distinguish between the context that ultimately supplies the demands on the system and the representations against which the system can be mechanically checked: *primary* versus *proxy* context.

A primary context is the source of a demand's authority. The compiler defines what programs it accepts, the runtime determines how the system actually behaves, and stakeholders decide whether the system meets their needs. A proxy context represents demands that come from a primary context. Specifications, pattern documents, templates, and rule catalogs borrow their authority from the corresponding primary context. Because that authority is borrowed, a proxy can silently diverge from its source. This can happen in three ordinary ways: the primary changes and the proxy does not; an update reaches some copies and not others; or the transcription was wrong from the start. The remedy is *reconciliation*: recurring checks that compare a proxy with its primary where the primary can be consulted mechanically, and that otherwise compare a ruling's proxies with one another (Table 2).

Not all primary demands have the same origin. Some are settled empirically: the compiler, the runtime, and the world behave as they behave, regardless of what any document says. Others are created by authoritative rulings. The primary here is the ruling itself, not the documents that record it. An architecture decision on the solution side, and a stakeholder ruling on the problem side, create those demands simply by declaring them. Primary versus proxy is therefore a distinction of authority, not of medium: a ruling may itself be written down, but the authoritative act is primary, while copies that derive their authority from it are proxies. A ruling's primacy makes it the standard for its proxies, not a fact about the world: whether the demands it creates still fit the world remains the outer loop's judgment (S = W). Table 2 labels the two kinds *primary (empirical)* and *primary (ruling)*. The distinction applies to both the solution and problem sides of the ensemble. Every context the generated system is measured against is either primary or proxy (Table 2).

*Table 2. The two-sided, two-mode structure of context, and the reconciliation process that spans it, as instantiated in the reference implementation.*

| | **Solution side** | **Problem side** |
|---|---|---|
| **Primary (empirical)** | Facts of the stack such as what the compiler accepts, how the framework binaries behave, and how aggregates are reconstructed by event replay, consulted by the harness through compilation, tests, and executable contracts | Facts of the world such as how time, concurrency, and identity actually work, consulted through demos, observation, and lived use |
| **Primary (ruling)** | Conventions the team creates by declaration, recorded in documents such as architecture decision records; violating them breaks nothing physically | Demands the stakeholders create by ruling, such as what a feature must do and when it counts as done; the specifications only transcribe them |
| **Proxy** | Documents that restate primary demands, such as pattern documents, templates, API references, and most deterministic-review rules | The specifications themselves, such as `frame.yaml`, requirements, the machine and domain specifications, and acceptance scenarios |
| **Reconciliation** | Mechanizable checks, such as imports against the framework jars, versions against manifests, and ports against a single designated configuration file at every session start, and dual-profile tests against the live framework at every gate run | Not mechanized: machines can gather and interpret the evidence from demos, stakeholder reviews, and lived use, but no machine judgment establishes that the proxy matches its primary |

Three practical consequences follow. First, the *false-green condition*: a form can fit the proxy while misfitting the primary, so all checks against the proxy can report green while the form remains wrong. The methodology's self-audit machinery catches this class of defects where proxy–primary reconciliation can be mechanized; the outer loop covers the remainder. Examples include a pattern document that has drifted — a proxy failure — and a stale rule checker or a gate that passes while checking nothing — a failure of the evaluating machinery, caught by the fixture and reachability checks of Section 3.4 rather than by reconciliation. A theory that does not distinguish proxy from primary cannot even express the first kind.

Second, *gate density follows the physics of enforcement*: gates are concentrated where misfits do not reveal themselves. The compiler automatically rejects invalid programs. The running system exposes some defects, but only through tests and the contract clauses they execute. At the design level, however, the platform itself neither prevents nor exposes violations of aggregate encapsulation or layering. Those violations therefore require explicit detectors, which is why the 188-rule catalog is concentrated there.

Third, *when a primary cannot be consulted directly, one proxy must serve as the designated original for mechanical reconciliation*. An empirical primary can be consulted directly, so its proxies are checked against the primary itself: imports against the framework jars and tests against the live framework. A ruling cannot be consulted this way: its proxies can still be checked against one another, but their agreement does not establish agreement with the ruling itself. This is why each rule has one designated original (Section 3.4).

Note that the specification *S* in the mnemonic chain *S* = *P* = *T* (specification = program = tests) is itself a proxy of the world *W*. We therefore name the complete structure the *S* = *P* = *T* = *W* *assurance model*. The name follows software engineering's own division: verification asks whether the program and its tests match the specification, validation asks whether the specification matches the world, and assurance requires both; the methodology reserves the validation judgment to humans (Section 4.3). The equals signs are mnemonic rather than literal: the model comprises three families of machine-decidable relations enclosed by a fourth relation that requires human judgment (formal semantics, including the converse check that nothing undeclared is realized, in Appendix A), a point to which Section 6 returns.

## 4.3 Detection cost sets the affordable feedback cadence

Distinguishing the two kinds of context says nothing about how often each can be consulted. In the vernacular, misfit detection was embedded in dwelling rather than performed as a separate activity, keeping the cost of each additional check low enough for continuous feedback. The Oregon Experiment instead institutionalized periodic diagnosis [9]. Diagnosis required human judgment, so it was too costly to perform continuously. The same constraint applies in both settings: *the cost of each additional check sets how often feedback can be afforded* (stated as an inequality in Appendix A).

The methodology makes this constraint explicit in its architecture. The inner loop automates the checks, so they can be run continuously at low marginal human cost. This allows feedback at machine speed, restoring the rapid corrective cadence that everyday use provided in the vernacular. The outer loop, however, ultimately requires human judgment, so its cadence must be budgeted. A fast review cycle runs at each delivery boundary, where stakeholders observe the system's behavior. A slower cycle runs during long-term use, asking whether the system still serves the people who live with it.

Machines can still help in the outer loop, but their judgments may not substitute for the human judgment that establishes whether the specification still fits the world. They can remind the team when no review has happened for too long, gather the evidence for that review, and keep every verdict recorded and traceable. The methodology reserves that judgment to humans; machines can check whether it has been made.

Together, these elements recreate the feedback structure of Alexander's homeostatic process in engineered form [4]. A vernacular settlement stayed in good fit because people could notice local misfits and correct them quickly. The dual-loop methodology reproduces this feedback structure in software. The inner loop repeatedly checks each generated candidate against the misfit list and drives repair until the detected misfits disappear. The outer loop compares the specification with the world through human judgment, and it also maintains the representations and detectors on which the inner loop's verdicts depend. Domain-driven decomposition bounds how far repairs propagate, supplying the locality this process requires (Table 1, P5), but locality alone is not enough. The process works only when manufactured tradition (Section 3.4), manufactured manifestation (Section 3.5), and mechanized repair (Section 3.6) work together. Alexander's homeostatic process therefore corresponds not to any single pattern but to the verification harness as a whole.

## 4.4 The severed backflow

The inner loop of Section 4.3 repairs generated artifacts. This section asks who is allowed to change the tradition itself. In the vernacular, the people who repaired buildings also changed the tradition, because they belonged to the community that carried it. Repair flowed back into the tradition. With a builder subject to the bias current, that backflow would be unsafe. A generator allowed to amend the tradition would expose the law itself to the same learned defaults that the law is meant to constrain.

The theory requires that the governed builder may not, on its own authority, amend the law that governs it. The methodology therefore assigns the two responsibilities to different actors. The machine repairs generated artifacts, but it may not amend the tradition. In this methodology, final legislative authority is reserved to humans, exercised through a deliberate legislative circuit. Every defect that escapes the gates must then be answered in one of exactly two ways: by bringing the failure class into the mechanized enumeration $M$ of Appendix A — through a new gate anchored in the law, with a low false-positive rate, or through a repair to an existing detector, its reachability, or the proxy it reads — or by an explicit ruling that places the failure class outside $M$ and names the means that cover it instead. The first outcome extends or corrects $M$; the second is recorded outside $M$, so no claim of machine-checked fit rests on the class it names.

The hybrid that results embeds the repair dynamic of the unselfconscious process, now mechanized, in a self-conscious legislature — a configuration, we believe, Alexander did not describe. It lets each process do what it does best while avoiding what it does poorly. The self-conscious process is overwhelmed when the interacting misfits grow too complex to treat as a whole [4], and the unselfconscious tradition adapts only at the pace of cultural change. In the hybrid, the repair work that would overwhelm a designer is mechanized in the inner loop, and the tradition can afford to change slowly, because routine artifact repair no longer waits for it.

## 4.5 The permanent novice

Section 4.1 described the builder at a single generation: biased and native to none. This section asks not whether the builder can learn the local tradition, but whether it can be shown to have internalized it well enough for the scaffold to be retired. *The Timeless Way* ends by saying that the builder should eventually leave the language behind (Section 2.3). At first, explicit patterns guide the builder. With experience, the builder comes to respond to the forces directly, and the scaffold is no longer needed [10]. Alexander's endpoint presupposes a builder who can eventually act without continuing dependence on the explicit language.

The methodology must treat the builder as a *permanent novice* in the local dialect, even though its underlying model is already an expert in the mainstream. The reason is not that it learns slowly. It is that observed compliance cannot establish the state Alexander's forgetting requires: that the builder can henceforth be trusted to act from the tradition without consulting or being checked against it.

Alexander's forgetting is a claim about a state: the builder no longer needs the language and responds to the forces directly. A generator offers no equivalent warrant for retiring the scaffold. We can check what it produced this time. We cannot check that it has absorbed the tradition rather than matched it once. Observing the generator's output does supply evidence: it can raise confidence that the learned defaults have shifted. It cannot supply the warrant that retirement requires: measured reliability is evidence about the builder, not a verdict about any particular artifact. P1 makes the same point categorically: machine-checked fit rests on detector verdicts over each artifact, never on generator confidence. Retiring enforcement on reliability evidence would not establish the artifact claim more cheaply; it would substitute a claim about the builder for the claim about the artifact. Two further facts push the same way. The tradition keeps moving through the legislative circuit of Section 3.7, so any internalization would be a snapshot of one moment. And compliance with stated constraints degrades as the constraints accumulate, a failure independently documented as constraint decay [22].

Alexander's gate is passed through and eventually left behind (Section 2.3). Here it must be passed at every generation. Toward the mainstream, the model behaves as if it had *already* completed the forgetting that Alexander hoped for. It writes fluent mainstream code without consulting explicit rules. Its learned defaults supply what the local tradition must supply explicitly.

A harness is itself a persistent external memory, which raises the obvious objection: if the builder reads the tradition from files at every generation, has it not remembered it? The answer separates memory from internalization. Alexander's forgetting means the builder no longer needs to consult the language. Our builder must consult it every time, and must then be checked against it. The memory is real, but it sits outside the builder and is enforced rather than possessed.

Fine-tuning is the strongest form of the objection, because it moves the dialect into the weights rather than keeping it in a file. It may reduce how much of the tradition must be supplied as guidance (Section 8), but it gives no grounds for retiring verification. A retrained model carries learned defaults as of one training run; evaluation can show that those defaults have shifted, not that any particular subsequent generation will comply with the tradition. A retrained builder is still a generator whose learned reliability does not establish the compliance of any particular output, so its output requires verification against the governing tradition (Section 4.1).

Humans, by contrast, can follow Alexander's original path. With repeated use, the specification language can become part of their ordinary practice instead of something they consciously consult. What internalization retires, however, is guidance, not verification. The machinery of Sections 3.4–3.7 is therefore not temporary scaffolding. It is a permanent institution: even evidence that the generator had completed the same learning process would not discharge the obligation to verify each artifact (P1).

Some obligations do not arise specifically from the builder being an LLM and therefore match long-standing practice (P1, P2, P4, P5, P7). Others take on their present force because no one can establish that an LLM builder has internalized the tradition, because it defaults to the mainstream, and because it can be guided only through representations (P3, P6, P9, P10). P8 sits between the two groups: any builder faces a residue of judgment, but mechanizing the builder forces that residue outside the autonomous inner loop. For the second group and for P8, that force follows from what the builder is, not from arbitrary methodological choices.

# 5. Empirical Evidence

The theory is instantiated in a single system. We do not claim that the architectural binding is universal. We claim only that one working system implements a mechanism for every obligation the theory derives. Section 8 explains what this single case does and does not show.

The reference implementation implements the core domain of a Scrum-management system (products, sprints, product backlog items, and Scrum teams). It is built with domain-driven design, Clean Architecture, event sourcing, CQRS, and design by contract on Spring Boot/Java, together with a React/TypeScript frontend. Generation and verification ran inside an agentic coding environment (Claude Code, Anthropic), with the harness implemented as a plugin that attaches the gates and hooks to the environment's extension points.

Executable contracts are enforced at runtime by the uContract library [23]. The library, like the rest of the supporting framework family, belongs to the local dialect rather than to the mainstream. The builder cannot be assumed to know these APIs, so the written tradition must supply them (Section 4.1).

Development began in July 2025 and remains ongoing. During this period, generation used Claude Sonnet, Opus, and Fable models, and each change of builder model was followed by a fresh rebuild. The system is not a re-implementation of existing software. Its specifications were developed from scratch through Event Storming and problem-frame analysis (Section 3.2). Its backend has been built, and repeatedly rebuilt, *from empty source directories* using only the methodology. All 64 problem-frame specifications across four aggregates were executed under the full gate stack.

The rebuilds also exercised the harness itself. Whenever the gates or patterns changed, we deleted all previously generated source code and regenerated the backend from scratch. Every rebuilt backend then had to pass the end-to-end test suite, which is maintained by hand and never regenerated with the backend, driven through the unchanged frontend. To the extent captured by its assertions, the suite served as a regression oracle held fixed across rebuilds, in the manner of a characterization test [24]: behavior accepted in one rebuild had to remain compatible in the next, so each new rebuild was checked not only against the specifications but also against the accepted behavior of earlier rebuilds. After a series of rebuilds was complete, we manually ran an endpoint-coverage check to confirm that the suite exercised every endpoint declared in the specifications. The backend grows one use case at a time, while these checks exercise the whole system. We therefore run them only after a complete rebuild, outside the fix loop.

The frontend itself was vibe-coded once, accepted through direct human validation, and was not regenerated thereafter. This division of assurance labor follows the economics of Section 4.3: an artifact generated once can afford human-priced validation, while the endlessly regenerated backend cannot. Table 3 summarizes the specification corpus, the written tradition and verification harness, the generated backend, and the rebuild outcomes. Every count of the tradition and the harness is a snapshot of an enumeration that is open by design (P8). Gates are added as escapes are adjudicated (Observation 1), and advisory gates are promoted to blocking gates once they have shown no observed false positives in use. A green rebuild requires not only zero reported violations but also evidence that every blocking gate actually ran: the gate runner counts a gate that could not run as skipped rather than passed, and its overall verdict separates all-passed from passed-with-skips, so a rebuild is accepted only when every blocking gate both ran and passed (a gate-level approximation to the $M_{\text{eval}} = M$ requirement of Appendix A).

One number in Table 3 needs explanation. The specifications total about 40,000 lines of YAML, which exceeds the roughly 24,000 lines of production code they generate and matches the size of the generated test suite. The composition rows in Table 3 show where those lines go, based on a line-level classification of the corpus. About half of those lines are formatting and recurring structure rather than new domain-semantic content.

The interface and data declarations record the attributes of aggregates, entities, and value objects and the fields of endpoints and queries. These declarations are numerous, but they record few new business decisions. Most restate decisions already made during Event Storming and recorded on the class diagram. Likewise, a use case in this architecture mostly wraps one aggregate command, so once the aggregate is defined, much of the use-case file, the contract clauses, and the initial acceptance scenarios follows from that definition.

*Table 3. Assets and outcomes of the reference implementation at the time of writing.*

| Asset / outcome | Measure |
|---|---|
| **Problem-side specifications** | |
| Problem-frame specifications (four aggregates) | 64 (≈500 YAML files, ≈40,000 lines) |
| — Interface and data declarations | ≈8,700 lines (21.7%) |
| — Acceptance scenarios (Given–When–Then) | ≈3,200 lines (7.9%); 310 scenarios |
| — Contracts (preconditions, postconditions, invariants) | ≈3,400 lines (8.4%); 547 declared clauses |
| — Behavioral rules (execution steps, state and reaction rules, error mapping) | ≈1,800 lines (4.6%) |
| — Requirements | ≈2,600 lines (6.5%) |
| — Recurring file structure and traceability metadata | ≈11,100 lines (27.6%) |
| — Blank lines and comments | ≈9,400 lines (23.3%) |
| **Written tradition and verification harness** | |
| Blocking gates | 28 |
| Deterministic-review rules | 188 |
| Specification-lint rules | 39 |
| Self-consistency checks | 21 |
| Architecture decision records | 53 |
| Pattern/doctrine documents | ≈145 (≈57,000 lines) |
| Scripts and their regression tests | ≈130,000 lines; more than 3,500 tests for the gate modules |
| **Generated backend (per rebuild)** | |
| Backend code | 336 production files (≈24,000 lines) and 201 test files (≈40,000 lines) |
| Test cases | ≈1,300, each run under both infrastructure profiles |
| **Rebuild outcomes and telemetry** | |
| Recorded deterministic-review runs and rule-level violations | 5,807 recorded runs (1 March–16 August 2026; the telemetry instrument was added partway through the project); 216,148 recorded violation reports at 1,748 distinct rule-and-file sites across 235 rules. The set of rules changed over time as rules were added and retired |
| Each aggregate's most recent rebuild | All blocking gates passed; all generated tests green under both profiles |
| PIT test strength [25] (killed / covered mutants; four aggregates) | 95–100% |

The 95 always-active rules of Section 3.4 are a subset of the 188 rules in the catalog. The 235 rules in the telemetry row of Table 3 are a historical union rather than a snapshot: the catalog gained and retired rules over the project.

The authors hand-wrote the first specification of each frame type, mainly its domain part, and revised it with the model. Once that structure had stabilized, the model generated the later specifications of the same type from the Event Storming board, the class diagram, and an existing example, and the authors reviewed and revised what it generated. This reduced the manual input.

The material that required the most author judgment in this case concentrates in about 27 percent of the corpus, roughly 11,000 lines: the acceptance scenarios, the contracts, the behavioral rules, and the requirements. The percentage locates the judgment-bearing material; it does not measure the effort behind it. Section 8 discusses what concentrating human judgment in this material achieves and what it does not.

A second number in Table 3 needs a caution. The PIT scores measure the sensitivity of the tests to the program mutations generated by PIT. They say nothing about whether the tests are faithful to the acceptance scenarios. That faithfulness is the S = T leg of the model, and Section 8 discusses it as the least directly verified of the three machine-decidable relations.

The telemetry row supports a first analysis, and its window must be stated: the instrument was added partway through the project, so the record covers 1 March to 16 August 2026. Most raw counts repeat, because every run re-reports a standing violation until it is repaired. Deduplicated by rule and file, the 216,148 recorded violation reports reduce to 1,748 distinct rule-and-file sites across 235 rules. The rules with the most distinct sites are consistent with the bias-current interpretation of Section 4.1. Six of the top eight guard the local test dialect, its reporting annotations, its feature declarations, and its requirement that every controller carry both a unit test and an integration test. The seventh guards constructor null-checking, and the eighth guards the audit identity that every command must carry. Each is a specific local requirement that generation repeatedly failed to preserve. These counts do not identify the cause of the violations; they show only where violations concentrated. Per-model violation profiles remain future work (Section 8).

The observations fall into three families. Together, they exercise several of the theory's propositions rather than evaluate it exhaustively.

**Observation 1 (P1 *negative fit*, P2 *manifestation*; P6 *tradition-bearer*, P8 *residue of judgment*). Locally well-formed artifacts can still misfit at their boundaries; boundary escapes exposed gaps in the enumeration.** The closeout of each rebuild performs a final verification pass over the whole aggregate. It repeatedly found defects that no per-artifact inspection had caught. Examples included an input field that the receiving service silently never reads, a controller error branch that can never execute and therefore collapses specified errors into HTTP 500s, a mapper that loses derived state when reconstructing an aggregate from storage, and a specified cross-aggregate reaction that was never wired. Each artifact is locally well-formed, so these defects remain invisible when inspected one artifact at a time. The misfit lies in the chain between them (Section 3.5).

Under the two-way answer of Section 4.4, every escape has exactly two permissible outcomes. It either extends or corrects the enumeration or receives an explicit ruling that places it outside $M$ and names the non-gate mechanism that covers it.

Examples of new gates include checks for input-field consumption chains, controller error-branch reachability, boundary-value contract tests, and value-object nullability. Semantic faithfulness to the specification's intent, beyond what the deterministic gates already capture, cannot be decided by the harness. The methodology therefore assigns it to the advisory multi-model review of Section 3.5, not to the deterministic fit predicate; the review's findings go to human adjudication (Section 3.6).

Quietly fixing the individual defect and moving on is not permitted. When the misfit list proves incomplete, human judgment re-enters through post-mortem analysis to extend the enumeration, as the residue-of-judgment concession of Section 2.1 predicts.

**Observation 2 (P2 *manifestation*, P10 *formal picture*). Proxies drift, and so do the detectors themselves.** Independent audits of the verification artifacts found the same failure class at multiple levels. Documentation reported counts that no longer matched the rules. Hooks still checked retired doctrine. Gates passed only because their scan set had silently narrowed. A validator's own docstring still counted a dead script as wired.

Two of these failures, the stale counts and the stale docstring, are examples of the proxy drift that produces the false-green condition of Section 4.2, in which the form fits a proxy that misrepresents its primary; the other two, the hooks checking retired doctrine and the narrowed scan sets, are failures of the evaluating machinery itself (Section 4.2). Each failure led to a new check. The artifacts that verify the code must themselves be verified. They are checked regularly against fixtures that show the rules still fire and against real invocations that show the rules are still reached.

**Observation 3 (P1 *negative fit*, P3 *viscosity*). Prompt-carried guidance was repeatedly violated; the corresponding gates caught the violations.** The project's own history showed this observation early on. The project began with prose alone: a large collection of prompts and sample code with no executable checks. With prose alone, the same classes of rule violations continued to recur.

We therefore added deterministic gates one by one. The model wrote each gate, but the authors decided what it should check. The written tradition evolved in the same way. The authors wrote the first core patterns by hand. After the system had stabilized, the model drafted later pattern documents from sample code supplied by the authors, who again decided what should be checked. This arrangement does not delegate normative authority to the model: the decision of what to check remains with the authors (Principle 6), and the detectors themselves are verified so that the implementation of that decision stays faithful (Observation 2).

We then substituted other models of the same family, while keeping the harness unchanged; these were working rebuilds rather than controlled comparisons (Section 8). Prompt-carried rules were violated repeatedly, but the corresponding gates caught those violations. The same pattern appeared in maintaining the proxies. Session-start reconciliation checks repeatedly detected proxy drift before generation began, for example by comparing pattern imports with framework jars, versions with manifests, and network ports with the designated configuration file. These observations are consistent with the theoretical claim that enforcement, not guidance, must carry compliance (Section 4.1).

Taken together, these observations show that the methodology can repeatedly regenerate implementations of the same specified system that remain compatible with the behavior exercised by the fixed end-to-end suite, while its verification machinery continues to evolve. The observations therefore exercise the theory in operation rather than on a benchmark.

Table 4 states the observation-to-theory correspondence.

*Table 4. Observations mapped to the theory they exercise: propositions (Table 1) and derived results (Section 4.2).*

| Observation | Theory exercised |
|---|---|
| 1. Locally well-formed artifacts can still misfit at their boundaries; boundary escapes exposed gaps in the enumeration | P1, P2 (manufactured detection); P6, P8 (the human legislature grows the open list) |
| 2. Proxies drift, and so do the detectors themselves | P2 (fixture-proven liveness and reachability); P10 (proxies under verification); the false-green condition of Section 4.2 |
| 3. Prompt-carried guidance was repeatedly violated; the corresponding gates caught the violations | P1 (machine-checked fit rests on detectors, not on the generator); P3 (tradition externalized *and* enforced); the mechanizable proxy reconciliation of Section 4.2 |

The remaining propositions are instantiated by the machinery but not separately measured as empirical observations. Every fix-loop run freezes its context (P4). Every aggregate rebuild operationalizes repair locality through bounded propagation and the regression guard (P5). The architecture gives the inner and outer loops different feedback cadences (P7). The same scaffold has remained in place throughout the project (P9). Measuring each of these individually — for example, how convergence behaves within a frozen context, how often repairs remain local, how frequently each loop runs in practice, or how often each rule still fires — remains future work.

This evidence has clear limits. It comes from one project, one architecture, one solution stack, and one primary legislator of the tradition, who was also the Scrum domain expert. As a result, the domain knowledge elicited during Event Storming, the resulting specifications, and the S = W judgments of the outer loop all came from the same person. The case therefore does not evaluate situations in which problem elicitation, specification authorship, and domain expertise are distributed across different people. It demonstrates feasibility and operational coherence, but not portability (Section 8).

# 6. Answering Alexander's Challenge

Alexander's OOPSLA address [5] posed a graded challenge: is your language generative; does it aim at wholeness; does it make life better?

**Generativity: answered operationally by supplying the missing executor.** In the three decades since patterns entered software, mainstream practice has used them mainly as reference literature: read and applied through professional judgment rather than executed as a written generative procedure. Deterministic generators, from templates to model-driven and grammar-based synthesis, could execute a pattern only after its operative content had been rendered as formal rules, with the prose and its rationale left as documentation the generator never read. The LLM and the harness together supply both at once: the model interprets a pattern language as written, and the harness governs its output deterministically. The governance is necessary precisely because interpretation does not guarantee compliance (Observation 3). Under it, the language demonstrably generates. A specification goes in, a structurally uniform system verified against the encoded obligations comes out, and the process is repeatable enough to rebuild a system from empty directories. Those thirty years were not wasted, however. They left a written tradition cataloged across scales, from problem classification [12] through architecture and design [14, 16, 17, 21] to acceptance scenarios [26].

A team manufacturing its tradition today curates a mature literature instead of first writing one. Alexander, in building, had to articulate his own from a tradition that was still largely tacit.

This execution model also reproduces an emphasis in Alexander's later work on *generative sequences of structure-preserving steps* [27, Book 2]. Each problem-frame type's workflow is an engineering analogue of such a sequence: a path through the pattern network of Section 3.3, unfolded layer by layer, with regression guards verifying that each transformation preserved the machine-checked properties previous steps established. The analogue is bounded: the guards preserve machine-checked properties, not Alexander's living structure.

**Coherence: answered in the formal sense, open in the experiential sense.** Aggregate-wide, cross-frame, and cross-aggregate gates check that the parts compose into a consistent whole, coherence in the formal sense. Alexander's *felt* wholeness [27] is a different property — structural in his account, not a matter of opinion — and we make no claim to have mechanized it.

We observe instead that when routine construction and repair are delegated to machines, as they are here, code is no longer the artifact that people routinely read, modify, and reason about. The level at which people work moves up, a shift consistent with recent spec-driven work [28]. On our account, people now inhabit the activities that frame and govern construction: discovering the problem, formalizing it into specifications, and maintaining the tradition that governs what the machine builds. Moving up does not make the code alien, however. The four forms confine the builder to the team's encoded tradition (Section 3.4), so what the machine writes remains within the team's architectural vernacular. When someone does step down to the code, to adjudicate an escape for example, they are reading code constrained to conform to their own encoded tradition. The claim is not that generated code is intrinsically easier to understand, which would require empirical study. The claim is that enforced tradition preserves a shared architectural vernacular across machine-generated artifacts, keeping the system culturally recognizable to the maintainers who will eventually inspect or modify it. In this sense, governance serves verification and preserves a precondition for habitability [29]. One form of the quality-without-a-name question (in software terms: is this a structure a person can inhabit, revise, and think in?) therefore arises most immediately at the frames rather than at the generated code. That question is open, and no gate should pretend to close it.

**The moral dimension: located but not mechanized.** Every machine-verified relation in the methodology relates artifacts to artifacts: code to specification, specification to schema, detector to fixture. The relation Alexander actually cares about (does the structure nourish the lives lived with it?) holds between the built thing and the world. Because the machine-checked legs already tie the program to its specification, the model carries what remains of that relation as $S = W$, specification against world. Section 4.2 shows why no machine verdict reaches it. The problem-side primary cannot be summoned into a test harness. The outer loop of Figure 2 reserves the slot for this judgment. The limit of this answer is that a slot is not yet an institution.

Alexander did not merely argue that human judgment matters. He also proposed institutional forms for exercising it. One is the Oregon Experiment's diagnosis [9]: periodic, formatted, pattern-anchored, budget-connected human judgment of the built world's health. Translated to software, the theory calls for an institution of this kind; we propose a scheduled inhabitation review as one such mechanism. Its findings would be recorded, traced to the frames they implicate, and fed back into amendments of the governing representations. Machines supply the reminders, the evidence, and the records (Section 4.3). Humans supply the verdicts. One participatory practice already exists at discovery time, the Event Storming workshop of Section 3.2; but at present, the participation of non-engineers is not carried forward into formalization.

The economics of Section 4.3 make the outer loop slower and more expensive than the inner one. The ethics of Alexander's OOPSLA address say the priority runs the other way: the inner loop exists to serve the outer. Making the three machine-decidable families of S = P = T = W routine (Appendix A) is not the goal in itself. Its purpose is to free human attention for the fourth relation, S = W, the one that requires human judgment. The engineer's work does not vanish in this division. It relocates, to the outer loop's contact with the world the software serves, and to the building of the tradition and harness themselves. Taken together, the three answers are consistent: mechanize what can be checked, relocate what cannot be checked, and do not let the machinery substitute for the judgment.

# 7. Related Work

**Alexander's software reception.** The software-patterns movement took shape through Beck and Cunningham's early adaptation of Alexander [30], the Gang of Four catalog [16], and the Pattern Languages of Program Design community [31]. Gabriel [29] kept questions of habitability, piecemeal growth, and Alexandrian generativity in view (questions that much of the catalog-oriented reception treated only weakly), and Section 6 follows that line of inquiry. In our view, this tradition generally lacked not insight but a *governable executor*. Patterns could teach their users, but their application still depended on human interpretation and judgment. Even when a language prescribed the order in which its patterns apply, that grammar was interpreted by its users rather than enforced and executed as a machine-verifiable generative process. The four-form obligation (Section 3.4) is what becomes of a pattern when its reader can no longer be trusted to exercise situated judgment consistently.

**Specification and contract traditions.** Jackson's problem frames [12] supply the classificatory basis of our problem representation; Meyer's design by contract [13] supplies its executable contract layer; and behavior-driven development [26] supplies the Given–When–Then scenario grammar that our acceptance specifications formalize and machine-check. Recent evidence suggests that making preconditions and postconditions explicit can improve the accuracy of LLM code generation [32]. Evans's domain-driven design [14] supplies the principal solution-side vocabulary, supplemented by Vernon's treatment of aggregates, domain events, and event-sourced implementation [19], and arranged within the layered macro-structure and dependency discipline of Martin's Clean Architecture [21]. Mutation testing [25, 33] serves, in our stack, as a meta-level adequacy check over executable verification, checking whether the tests, and the contract clauses they execute, detect deliberately injected defects. Taken together, these traditions contribute problem classification, executable contracts, acceptance scenarios, architectural vocabulary and structure, and tests of verification adequacy. None of them, as conventionally combined, provides a criterion for when an AI-coding verification harness is sufficient to justify a stated claim of fit. This is the question addressed by the paper's misfit calculus (P1; Appendix A).

Recent specification-driven work on AI-assisted development [34] goes further by formalizing specification governance as a reference model. In this model, structured specifications serve as the authoritative source of truth, stochastic generation must pass deterministic validation, and generation, verification, and governance close into a loop. This specification-centered approach has an industrial counterpart in Spec Kit [35], which, in the version examined here, externalizes intent into persistent, structured artifacts rather than leaving it only in conversational prompts: a constitution declares project principles, an ordered path runs from specification through plan and tasks to implementation, cross-artifact analysis checks those representations for consistency, and a convergence step checks the resulting code against them. Both lines structure the path from specification to implementation. Our theory addresses a logically prior question of sufficiency: what must the verification boundary around such a path establish before the absence of detected misfits can justify a claim of fit? The question remains because the specification and the detectors that enforce it may themselves misrepresent or drift from their sources. A claim of fit is justified only relative to the misfits that have been explicitly identified and checked. The primary–proxy distinction makes the harness's own failure modes expressible, and the legislative circuit governs their repair. Read against Table 1, a workflow of this kind could be bound into an implementation of the theory; the table states the obligations it would still have to discharge, among them which constitutional constraints must be operationalized as detectors, how the manifestation of each declared misfit is manufactured and its detector kept live, and how amendments to the law are kept outside the autonomous fix loop.

The sufficiency question has a close neighbor in safety-critical systems. Assurance cases justify critical claims through structured arguments over evidence [36], and eliminative approaches build confidence by identifying and resolving defeaters, the ways in which those claims or their supporting arguments could fail [37]. This negative logic is closely related to misfit enumeration. Like misfit enumeration, eliminative induction makes confidence conditional on the identified ways in which a claim could fail rather than on an absolute proof that none remain. The difference lies in what that logic must become in a generative workflow. In Assurance 2.0 and eliminative approaches, defeaters remain part of an assurance argument to be assessed. The misfit calculus requires each misfit in the mechanized enumeration to be embodied in a deterministic detector whose liveness is itself verified, so that adjudication can occur at the rate at which the builder produces candidates (Section 4.3).

**LLM code generation, repair, and guardrails.** Code-generation benchmarks such as HumanEval [38] evaluate function-level code synthesis, while repository-level benchmarks such as SWE-bench [39] evaluate whether systems can resolve real-world GitHub issues in existing codebases. SWE-agent [40] shifts attention from the model alone to the design of the agent's working environment, showing that an agent-computer interface can substantially affect coding performance. These works study coding-task completion under an externally supplied evaluation oracle, whether generating a function or repairing a repository. Our theory instead addresses generating systems from explicit specifications, where the oracle is part of the problem rather than given in advance (Contribution 1). The evaluation target differs as well: these benchmarks ask whether the code completes the task, while fit asks whether the candidate exhibits any enumerated misfit — functional failure is only one class among them (Section 3.5).

Self-repair and self-debugging systems [41, 42] mechanize iterative correction using failing tests, execution feedback, or the model's own explanations of its generated code. They study how such feedback enables repair, including how self-generated explanation can support debugging even without external feedback on code correctness [42] and how feedback quality limits self-repair [41]. They do not ask how a verification harness should be constituted, kept live and reconciled with its sources, and amended when an escape reveals that the enumeration is incomplete (P1, P2, P6, P10). Once a gate has detected a manifested misfit, these mechanisms are compatible with our theory as repair strategies inside the fix loop (Section 3.6). They can make repair more effective, but they do not replace the gate stack that determines which misfits are made manifest and detected in the first place.

Reflective agents [43] let a model learn from feedback on its own attempts within a task, storing verbal reflections to improve subsequent trials. Their feedback loop serves task performance. Our theory instead institutionalizes the outer loop, in which humans judge whether the specification itself still fits the world (S = W; P7, P8). A related line of work moves behavioral constraints outside the model altogether. NeMo Guardrails [44] expresses programmable rails around LLM applications. AgentSpec [45] provides a domain-specific language for specifying and enforcing runtime constraints on agent behavior. ContextCov [46] transforms operationalizable constraints in agent instruction files into executable checks. These systems reinforce the premise that natural-language instructions alone do not reliably ensure compliance. Where constraints can be operationalized, executable enforcement makes compliance more directly checkable than instruction alone. Our question is one of sufficiency. We ask what makes a set of constraints sufficient to justify a claim of fit, how the proxies and detectors that encode those constraints are kept live and reconciled with their sources, and who may amend them when an escape reveals that the enumeration is incomplete. Across these approaches, the same residual risk remains. A generated artifact can pass every existing check while still violating requirements that no check covers. Our theory does not eliminate this risk. Fit is never claimed absolutely but only against a stated enumeration (P1, P8; Appendix A), and every escape must be answered, by extending or correcting the enumeration or by an explicit ruling (Section 4.4).

**Repository-scale coherence and specification-driven verification.** Four recent works approach parts of this territory from complementary directions. Structured spec-driven engineering [28] argues that structured specifications can guide repository-level generation while making its results easier to verify. Constraint decay [22] shows that coding agents become less reliable as structural requirements accumulate while the API contract is held fixed. This independently documents a failure pattern that the non-native builder model predicts (Sections 4.1 and 4.5). The patchwork problem [47] names a failure class our closeout gates also target: code that is locally valid but globally incoherent. It detects this failure deterministically by checking consistency invariants over graphs derived from the repository's own artifacts. It detects rather than repairs. CodeSpec [48] has an LLM build repository-grounded functional chains from feature requirements and repository evidence, compile them into executable architecture and behavior specifications, and check the implementation against them. Taken individually, these mechanisms are therefore no longer unique to any single system. Three differences remain. Our chains originate in pre-existing specifications rather than in the repository's own structure, so the checks compare the code with declared meaning, not only with itself [47]. Our specification language is law that the builder may not amend, whereas a specification the generator derives for itself [48] remains open to the generator's own bias current (Section 4.4). And our field gates trace each field the specification declares, checking name, type, and genuine carriage at every representation it crosses (Section 3.5).

Beyond particular interfaces and enforcement mechanisms, recent work increasingly treats the harness itself as a first-class engineering object [49]. One line catalogs the runtime responsibilities a harness must cover around a foundation-model agent [2]. Another evolves harnesses automatically against observed failures [3]. A third asks what makes a system a harness at all, giving the category an operational membership test [50]. A parallel strand engineers the process that drives the agent, not only the software it produces. One line of work formalizes that process as an explicit, written loop specification handed to the agent harness [51]. Another demands proof before lifecycle progress: each transition requires fresh, state-bound, machine-checkable evidence, and unsupported claims are not admitted [52]. Our contribution is complementary but distinct. Existing work has shown that agent performance depends on interface design, that natural-language constraints can be externalized into executable enforcement, that harness responsibilities can be cataloged and harnesses can be automatically evolved, and that the agent's loop can itself be specified and evidence-gated. We derive instead the obligations of a verification harness from Alexander's theory of form. The resulting sufficiency claim is relative to the mechanized misfit enumeration of Appendix A. The question is not merely which runtime components are useful, but what must be established before passing every existing check can justify a claim of fit. Enforcement alone is not enough: the detectors must remain live, the proxies they check must remain reconciled with their sources, the repair of detected misfits must be mechanized, and the enumeration itself must be governed. Automatic harness evolution [3] also sharpens a governance question our theory makes explicit: when the harness rewrites itself, is it improving a mechanism or amending the norm that defines correctness? P6 requires that the second kind of change pass through the human-gated legislative circuit.

**Design theory.** Simon characterized an artifact as an interface between an inner environment (its substance and internal organization) and an outer environment (the surroundings in which it must function) [53]. This scheme is structurally resonant with Alexander's form–context ensemble. The information-systems tradition later articulated *design theory* more precisely. Walls, Widmeyer, and El Sawy defined information-systems design theory as prescriptive and grounded design prescriptions in kernel theories [7], and Gregor and Jones distilled design theory into an eight-component anatomy [6]. Table 5 maps our theory onto all eight components of that anatomy.

*Table 5. Our theory mapped to Gregor and Jones's eight-component anatomy [6].*

| Component | Supplied by |
| --- | --- |
| Purpose and scope | Section 1 |
| Constructs | Misfit; non-native vernacular builder and its bias current; primary and proxy context; four-form pattern language; gate stack; dual loop; the S = P = T = W model (Sections 2–4) |
| Principles of form and function | The seven governing principles (Section 3); Table 1 |
| Artifact mutability | The generated artifact is mutable through the fix and outer loops, whereas the governing tradition is mutable only through a separate human-gated legislative circuit (Sections 3.1, 3.6, 3.7, and 4.4) |
| Testable propositions | Enforcement-over-guidance, false-green, and detection-cost predictions (Sections 4.1–4.3); portability and internalization conjectures (Section 8), with initial empirical evaluation in Section 5 |
| Justificatory knowledge | Alexander's theory of form [4, 5, 9–11, 27] as the principal kernel theory; Jackson's problem frames [12] as the principal problem-side foundation |
| Principles of implementation | Sections 3.2–3.7 |
| Expository instantiation | The reference implementation (Section 5) |

Finally, Naur's "programming as theory building" [54] anticipates a distinction central to this paper: a program's source code does not capture the understanding of how it relates to its world and why it is structured as it is. We extend the distinction by treating responsibility for the S = W judgment as non-delegable to the machine. Polanyi's account of tacit knowing [55] and the Dreyfus model's progression from rule-dependent novice to intuitive expertise [56] help explain the human process of internalization. Section 4.5 contrasts that process with the LLM: observed compliance cannot establish that it has internalized the tradition, and it must be given the tradition at every generation. To our knowledge, no prior work has derived the engineering obligations of an AI-coding verification harness proposition by proposition from Alexander's fit–misfit account. None has modeled the LLM as a non-native vernacular builder whose persistent pull toward mainstream idiom requires the local tradition to be enforced, violations to be made mechanically detectable, and the builder to be prevented from changing the rules by which its own output is judged — respectively, enforced tradition, manufactured manifestation, and severed backflow.

# 8. Limitations and Future Work

**External validity.** Section 5 noted that all evidence comes from one project, one architecture, one solution stack, and one primary legislator of the tradition. The theory's portability claim is precise and falsifiable, and it prescribes its own test. A second MGD binding holds the problem fixed (its stakeholder demands, classified by the same frame types; as Section 3.2 notes, the problem representation is architecture-dependent and may change), rebinds the solution side to a different architecture, technology stack, or implementation language, and asks whether the same obligations can again be discharged by different mechanisms. The present problem-frame artifacts separate the problem side from the solution side only in part. Gates already keep each frame's `frame.yaml` from naming any solution file outside the argument it records for each frame concern, and require the frame's requirements to be written in the vocabulary of the world, while a separate binding file, outside the five kinds of artifact listed in Section 3.2, names the architecture-dependent specifications (its machine and domain specifications) that realize the frame; those arguments and the acceptance scenarios, however, still cite solution-side artifacts and use the solution family's vocabulary. Completing the separation would confine solution-dependent changes to those specifications, making "holding the problem fixed" a property the harness can check rather than a methodological intention. Such a binding falsifies the theory if an obligation can be shown to be undischargeable in a binding that otherwise meets the theory's premises, or if the conditions from which the theory derives an obligation hold and the obligation can be removed without loss of the property it was derived to protect. This is not the only route: the derivation can also fail on logical grounds, if an obligation does not follow from its premises (Section 2.5). Because the derivation fixes the obligations but not the mechanisms, the failure of a particular implementation does not by itself count against the obligation it was meant to discharge.

A second binding may vary the implementation technology, the architecture, or both; varying the architecture reaches deeper. The evaluation fixes one solution-side family (domain-driven design, Clean Architecture, CQRS, and event sourcing), but Table 1 lists such choices only among the mechanisms, never among the obligations. A different family, say a CRUD-oriented three-tier architecture, would keep the frame types, which classify the problem rather than the solution, but would need its own specification schemas, decomposition units, pattern network, and gate set, while on the theory's account the obligations stand. Like the portability claim, this prediction is untested. We evaluated this family because it is the tradition we have built on since 2018 (Section 3.3) and because it is a constraint-rich case: its aggregate boundaries, inward-only dependencies, separate command and query models, and event-sourced state give the pattern network more to encode and the gates more to enforce. The theory prescribes how fit to a chosen context must be verified, not what that context must contain; a project that excludes architectural structure from its context thereby removes rules from its law, not obligations from the theory.

The obligations of Table 1 were derived from the classical theory plus one premise about the builder, so their claim to generality stands or falls with the derivation and its premises, not with the case. The case shows that the obligations can be jointly discharged in at least one working system: feasibility, not portability. Replication is therefore needed to test whether the obligations can be discharged elsewhere and how far the premises hold empirically, not whether the derivation is logically valid. The second-binding experiment is stated precisely so that any group can run it without us; we also intend to run it ourselves on a second architecture and technology stack.

**Builder variation.** The rebuilds were not designed to isolate the effect of model capability from that of the harness; they exercised the same obligations under successive builders. So long as the builder premise holds, changing the builder may change the violation profile, but it does not thereby change the obligations, which attach to the builder premise rather than to any particular model. Observation 3 shows that the violations were not peculiar to one model or model version, though all observed builders came from one model family; rebuilds with hosted and locally run LLMs from other families are the natural next test of that claim. How violation profiles differ across builders is a further question. The deterministic-review telemetry records every violation by rule, with severity, file, and timestamp, and Section 5 reports a first analysis. Counted per rule and per model version and normalized by each rule's evaluation opportunities, this record would map how the bias current manifests under a fixed law and harness, and would make a model swap under a fixed problem set and generation protocol auditable: a measured shift rather than an unverified assumption. Fine-tuning a model on the pattern documents is one more builder change to measure this way, and a possible mechanical analogue of Alexander's re-internalization (Section 2.3; Table 1, P9). We conjecture that it would thin the prompted guidance; Section 4.5 explains why it gives no grounds for retiring the gates.

**Of the three machine-decidable relations, S = T is the least directly verified.** The LLM currently translates each acceptance scenario into a test, and machines check the result only structurally: every scenario not deferred by the specification must have a corresponding test whose assertions match the kind of outcome its then-clause declares, and a state or quoted value the clause names must appear in that scenario's test method. No machine checks whether the test verifies what the scenario actually meant. The program is likewise generated by the LLM from the specification, but it is checked against the specification twice, directly by the structural gates of S = P and indirectly by running the tests; the tests are checked only structurally, yet they are the oracle by which the program's behavior is judged (Appendix A.2). A misreading of the specification that the model carries into both the program and its tests can therefore pass unnoticed, unless it also breaks a structural agreement that the S = P gates check. Mutation scores do not help here: they measure whether the tests detect changes in the program, a relation between tests and program only, and a test suite can reach a high, even complete, mutation score under the chosen operators while asserting the wrong business rule. The one mechanism aimed directly at semantic faithfulness, the multi-model review of Section 3.5, remained advisory. This acknowledged residual risk is addressed directly only by a mechanism whose effect has not been measured. Placing that review outside the misfit set $M$ (Appendix A.1) keeps this honest, since no claim of fit rests on it, but the gap in coverage is real.

We see three remedies, ordered from cheap to structural. First, trace every expected value a then-clause declares to an assertion of the corresponding test. A weaker form already blocks: a named state or quoted value must appear somewhere in the scenario's test method, while numbers and dates are compared only as warnings. Requiring each value to reach an assertion extends the field gates of Section 3.5 one hop further, from the data transfer object to the assertion; it cannot show that the assertion constrains the right subject or expresses the intended relation, but it catches an expected value that reaches no assertion at all. Second, mutate the specification, the mirror image of program mutation: perturb one expected value, regenerate only the test, and run it against the unchanged program; the test should now fail. While tests are generated by an LLM, a surviving mutant may reveal a clause that no assertion constrains or mere variation in the regenerated test, and a verdict this irreproducible can only advise (Section 3.5). Third, make the translation deterministic. A translator that derives each test, assertions included, from the acceptance specification turns the S = T leg from "generated then approximately checked" into "mechanically derived." Trust then concentrates in the translator, which must be verified like the rest of the harness (P10; Section 4.2), and the specification must express expected values formally enough to compile. The difficulty thus moves into a form the methodology already addresses through misfit enumeration and the specification linter. Once the third is in place and the mutation operators are restricted to mechanically valid perturbations whose expected outcome contradicts the original, so that no program can satisfy both the scenario and its mutant, the second can be promoted from advice to a gate.

**Appendix A demands more than the implementation uniformly delivers.** The gate runner enforces $M_{\mathrm{eval}} = M$ at gate granularity: a gate that cannot run is counted as skipped, never as passed. Within a gate, however, a detector whose scan set silently narrows can still pass without evaluating everything it should, and vacuously if the set is empty. The audits of Observation 2 caught this class, and several gates now fail loudly on an empty scan set, but uniform reporting of $|M \setminus M_{\mathrm{eval}}|$ at predicate granularity remains future engineering.

**The blocking gates enclose only the rebuilt backend.** The React frontend and the backend each hold their own picture of the HTTP contract, and no blocking detector compared their fields. Field names, date formats, and enum vocabularies drifted apart while unit tests and static types on both sides stayed green: two proxies of one truth, each checked only against itself, the class of failure that Section 4.2 predicts. Only end-to-end execution made the drift visible. A deep check that compares the specification, the backend controllers, and the frontend calls field by field exists but runs only as an opt-in audit; a shallower check, that the endpoints the frontend calls exist at all, runs as an edit-time hook. Extending the enumeration across this boundary is one direction in which the misfit list is currently growing.

**The economics of externalization.** We measured neither the effort of building the harness nor that of authoring the specifications. The theory expects the harness effort to be substantial: the harness carries the portion of the tradition that the builder cannot be assumed to know, and Section 4.5 argues that this scaffold cannot come down. Two economies bound the harness cost. The tradition needs to encode chiefly what the builder's learned defaults do not reliably supply (Section 4.1), although what to verify remains governed by the enumeration, not by the builder's reliability; and the frame concerns are amortized across every problem of a class (Section 2.4). Whether the cost pays for itself, within one project or across projects on the same stack, is unmeasured, and the deterministic-review telemetry above measures only interception, one input to the benefit side. Specification authoring is a cost paid per use case; it replaces prompt-writing and routine code review, and its judgment-bearing material is roughly the quarter of the corpus that Section 5 locates. The theory predicts not that this cost is small, only that it buys what prompt effort alone cannot: an externalized specification can anchor independent machine verification. A purpose-built specification editor could keep the class diagram drawn during Event Storming and the specification files synchronized in both directions, so that a developer would never need to edit the YAML directly. Because the price amortizes over repeated regeneration, it buys little in exploratory prototypes, in one-off tools, or while the architecture itself is still being discovered; the reference implementation draws this boundary itself, governing and regenerating its backend while its frontend was vibe-coded once (Section 5).

A third cost is the context window. We cannot yet say how much content the guidance forms (doctrine, template, and prohibitions), and the doctrine in particular, should carry: content beyond what correctness needs competes for the context that generation itself uses. In one informal comparison, a student group working by mob programming hand-wrote the guidance forms for one layer; generation worked under them, and they came to about half the length of their model-drafted counterparts but took far longer to write. Our working explanation is that a human author recognizes better which local choices the builder's learned defaults do not supply (Section 4.1). In this project the boundary between needed and excess content was found by trial, the most time-consuming part of forming the pattern language, and no principle yet predicts it. The gates face no such limit, because they run outside the generation context: one more asymmetry between guidance and enforcement.

**The outer loop as institution.** Section 6 called for institutions for the human outer loop; none has been built yet. Three pieces of design work remain: scheduled inhabitation reviews, in which humans periodically judge whether the system still serves the people who live with it; a field in each frame specification recording whose work the frame is supposed to improve, so far drafted for only a few frames, so that an inhabitation review can later ask whether it actually did; and the participation of non-engineers in authoring frames. This is where the methodology must grow if it is to answer the moral dimension of Alexander's challenge, not only its generativity and coherence.

# 9. Conclusion

Alexander's 1964 account [4] implies that, to the extent that the relevant misfits can be enumerated and mechanically detected, synthesis reduces to their technical elimination. That implication has become operational with the arrival of a builder productive enough to be worth governing, yet not to be trusted without verification. This paper has derived the harness's obligations from the classical theory, proposition by proposition, under the changed builder premise (Table 1). Under that premise, the broken enabling conditions of the vernacular become explicit engineering obligations. In this binding, misfit enumeration becomes a gate stack, tradition becomes a four-form pattern language, and vernacular repair becomes a regression-guarded fix loop. The methodology that results is Misfit-Governed Development: the builder proposes candidate forms, and the enumerated misfits govern what may be accepted. In this view, harness engineering is not a collection of ad hoc techniques but a discipline whose obligations follow from a single account of fit. One working system shows that these obligations can be jointly discharged; whether they can be discharged in other bindings is the test of Section 8. The residue of judgment (P8) stays where the derived theory places it: in the human outer loop, which judges whether the specification fits the world and approves every extension and amendment of the misfit enumeration. The machine's inner loop can be made routine and uneventful, and this is the point. Making it routine frees human attention for the outer loop, where Alexander's question remains: whether the software ultimately serves human life.

# Availability

The supporting libraries used by the reference implementation are open source at https://gitlab.com/TeddyChen. The ezddd libraries carry the event-sourced domain-driven design and CQRS machinery of the solution-side language (Section 3.3). The uContract and ezspec libraries carry the executable contracts and the Given–When–Then scenarios of the problem-frame specifications (Section 3.2). The reference implementation, its problem-frame specifications, and its verification artifacts (gate scripts, the deterministic-review rule catalog, and the pattern and doctrine documents) are not publicly released. The obligations themselves, however, are stated in full in the paper (Table 1), and the second-binding experiment of Section 8 is specified at the level of those obligations rather than of implementation artifacts. Independent replication would therefore require a new implementation of the stated obligations rather than reuse of our project-specific artifacts; a group running the second-binding experiment would hold a problem of its own choosing fixed across its two bindings.

# Acknowledgments

The work of Yu Chin Cheng was supported by the National Science and Technology Council, Taiwan, under Grant NSTC 115-2221-E-027-087-MY3.

# Competing Interests

Chien-Tsun Chen is the founder of Teddysoft, which offers training and consulting in domain-driven design, Clean Architecture, and AI-assisted software development. The open-source libraries used in the reference implementation (see Availability) are maintained under Chien-Tsun Chen's GitLab account and published under Teddysoft's package namespace. Yu Chin Cheng declares no competing interests.

# Generative AI Use Statement

The theoretical framework and methodology were developed by the authors. As Section 5 describes, Claude models (Anthropic) generated the reference implementation and, under the authors' direction, generated later problem-frame specifications, wrote the gate scripts, and drafted later pattern documents; Claude, Gemini (Google), and ChatGPT (OpenAI) served in the optional multi-model review (Section 3.5). Claude also wrote the scripts that computed the telemetry and corpus-composition figures of Table 3. Claude and ChatGPT additionally provided feedback on manuscript drafts and English-language editing. All theoretical claims, methodological decisions, interpretations of prior work and of the reported results, and final editorial judgments are solely the responsibility of the authors.

# Appendix A. A Minimal Formal Vocabulary

We fix a minimal formal vocabulary for what gains precision by it: the misfit calculus, the semantics of the S = P = T = W model, and the detection-cost bound. The vernacular and homeostasis material of Sections 2 and 4 is deliberately left in prose; formalizing it would add notation without adding precision.

## A.1 Misfits and fit

Adapting the formal apparatus of *Notes* [4] to this setting, let $G$ be the governing representation that a fix-loop window holds fixed: the specifications and the tradition against which candidates are mechanically checked (P4). $G$ is a proxy context in the sense of Section 4.2; the world $W$ stays outside it. Let $A$ be the set of candidate artifact states $a$, each comprising a generated form and its generated tests. The governing representation and the permanent verification artifacts are not part of $a$: their validity is a meta-level admissibility condition, stated below. *Notes* summarizes the state of each potential misfit by a binary variable; for a fixed $G$, we carry each mechanized misfit into the calculus as a *misfit predicate*: a decidable predicate $m_i^G: A \to \{0,1\}$, where $m_i^G(a) = 1$ means that artifact state $a$ exhibits misfit $i$ against $G$. The enumeration is the finite set $M = \{m_1^G, \ldots, m_n^G\}$: each $m_i^G$ formalizes one mechanized misfit class of the law and is evaluated by a blocking checker in the harness. Fit is the derived, negative notion:

$$\mathrm{fit}_{M,G}(a) \Leftrightarrow \forall m \in M: m(a) = 0.$$

Making $G$ a parameter rather than part of the artifact state records P4 formally: within a fix-loop window, successive candidates are judged against the same target, which is what makes one-at-a-time repair possible (Section 3.6); any amendment to $G$ ends the window and re-enters through the legislature of Section 3.7.

Two properties follow immediately from the definition; the third is an operational qualification. First, fit is *relative to the enumeration*: $\mathrm{fit}_{M,G}(a)$ asserts nothing about predicates outside $M$; this is the formal boundedness relevant to P8, and the reason the enumeration remains open. Second, fit is *decidable because every $m_i^G$ is*: the autonomous inner loop is possible because each $m_i^G$ was engineered to be a decidable predicate (P2); in the vernacular, physical consequences instead made many misfits directly manifest. Third, fit is *relative to the subset actually evaluated*: let $M_{\mathrm{eval}} \subseteq M$ be the misfit predicates a given run successfully evaluates. Each $m_i^G$ is total; the detector that evaluates it is not. A detector unable to resolve its subject — a scan set that matched no file or silently missed some, a specification in a format it does not parse — leaves its predicate unevaluated rather than evaluated to $0$. What a run establishes is $\mathrm{fit}_{M_{\mathrm{eval}},G}(a)$; the claim $\mathrm{fit}_{M,G}(a)$ is justified only when $M_{\mathrm{eval}} = M$, so a harness that reports fit without also reporting $|M \backslash M_{\mathrm{eval}}|$ claims more than it checked. The first property says the enumeration may be incomplete; this one says it may be incompletely applied, and the difference matters because a report that counts only the verdicts returned cannot show it. Recognition and mechanization are also distinct: the humans who govern the harness can recognize a misfit class, and rule on it under the two-way answer of Section 4.4, while the class still admits no decidable predicate; $M$ contains only the mechanized classes. The optional multi-model review of Section 3.5 deliberately sits outside $M$: it probes aspects of semantic faithfulness for which the methodology has no deterministic decision predicate. The advisory gates of Section 3.5 likewise sit outside $M$: their warnings inform the human legislature but do not enter the fit predicate.

Two boundary notes place the calculus. First, $G$ and $M$ together correspond to what Section 3 calls the law: $G$ is its representational part — including the records of rulings on classes not mechanized — and $M$ formalizes its mechanized misfit enumeration. Second, $\mathrm{fit}_{M,G}$ is an object-level predicate on candidates. P10 requires two kinds of checks on the representations: checks on $G$ itself (the machine-checked specification syntax and the partially machine-checked semantics of Section 3.2) and checks that reconcile proxies with their primaries (Section 4.2). These checks, together with the determinism of every checker (Section 3.5) and the fixture and reachability checks that show each detector still fires and is still reached (P2, P10; Sections 3.4 and 5), are meta-level admissibility conditions on the representations and detectors that make the predicate operational: preconditions for trusting a reported verdict, not additional conjuncts in the definition of fit. This reconciliation chain does not recurse indefinitely: it terminates at a mechanically consultable primary where one exists, and otherwise at the human ruling that authorizes the proxy (Section 4.2).

Sufficiency in this theory is therefore claim-relative: a harness is sufficient not when it establishes absolute correctness, but when it discharges every obligation needed to justify the bounded claim $\mathrm{fit}_{M,G}(a)$, including the evidence that $M_{\mathrm{eval}} = M$.

## A.2 The semantics of S = P = T = W

Let $S$ be the specification, a finite set of declared elements forming the machine-readable core of the governing representation $G$ of A.1; the relations below range over its fields, scenarios, contract clauses, error mappings, and the other deliverables it names (such as use cases, endpoints, and reactions), and the converse direction below also over its declared operations and domain events. Let $P$ be the program and $T$ the test suite (together, an artifact state $a$ of A.1), and $W$ the world. The equality signs are mnemonic: each denotes an agreement relation defined below, not mathematical identity between $S$, $P$, $T$, and $W$. This $S$ is the specification in the verification sense: the declared reference against which the first two families of relations below (S = P and S = T) check $P$ and $T$. The specification in Jackson's analysis [12] is narrower: it describes only the machine's behavior at its interface with the problem world and, together with the properties of that world, serves as a premise from which the requirement is argued to follow. Our $S$ also carries the domain specifications of Section 3.2 and the acceptance scenarios traced to the requirements. Read informally as three overlapping regions (the elements the specification declares, the program implements, and the tests exercise), the mnemonic asks that the three coincide over what $S$ governs: every declared element must be realized in the program and checked by the relation that applies to it, and nothing may be implemented or tested as a requirement that $S$ does not declare, since doing so would add a requirement no human has ruled on, amending the law (Principle 6). Which elements count as governed is fixed by the law rather than judged case by case: the specification schemas determine the kinds of element $S$ declares (Section 3.2). Everything else, whether supplied by the templates (Section 3.4) or written inside a declared element to realize it, is implementation or verification machinery that adds no governed element, such as mappers, adapters, private helpers, test fixtures, and test builders, and lies outside this comparison. The first two families below enforce the S-to-P and S-to-T directions, and the third requires the resulting tests to pass; the converse direction is blocking only where the families below say so and otherwise stays outside $M$. The machine-decidable portion of the S = P = T = W model comprises three families of relations:

*Structural conformance* (S = P): for every structurally declared element $e \in S$, its realization set $\delta(e)$ of program elements is non-empty and $\text{agree}(e, e')$ holds for every $e' \in \delta(e)$ along $e$'s journey, where agree is the family of agreement predicates of Section 3.5 (for fields: name, type, and genuine carriage; for error mappings: the controller gates' checks; for other named deliverables, at least presence). The converse direction (nothing realized that is not declared) is blocking for domain-event records, where an undeclared field enters the permanent log, and for the aggregate root's public behavior methods and the events each applies directly; for the other representations it is an advisory diagnostic and, per A.1, stays outside the fit predicate. This family carries the portion of S = P that direct structural inspection can decide; conformance of the program's behavior to $S$ is evidenced only indirectly, through the two families that follow.

*Scenario coverage* (S = T): for every acceptance scenario $s \in S$ not deferred by the specification (a specification may defer the tests of an entire test tier, such as its integration tests), its test set $\tau(s) \subseteq T$ is non-empty and each test's assertions are checked against the scenario's classified then-clauses; declared contract clauses receive their own generated contract tests. Conversely, a test that cites a scenario $S$ does not declare is blocking; what a correctly cited test additionally asserts is not checked. Faithfulness beyond this structural matching is a stated limitation (Section 8).

*Execution* ($T(P) = \text{pass}$): $P$ passes every test in $T$ under every configured infrastructure profile. These three families form the part of $M$ that compares the artifact with $S$; the rest of $M$ compares it with the tradition in $G$, which the mnemonic does not name (Section 3.4).

S = W is deliberately not a machine-decidable relation in the methodology. $W$ is the primary problem context of Section 4.2: its demands are not available to the harness as a complete machine-verifiable representation, but are encountered through lived use and stakeholder judgment. Formally, the model is therefore *three families of machine-decidable relations enclosed by one judgment reserved by construction to humans*. The design theory's economic claim, whose cost basis A.3 formalizes, is that routinely discharging the three machine-decidable relations makes sustained human attention to the fourth affordable.

## A.3 The detection-cost bound

For a detector $d$, let $c_d$ be the cost of one detection pass, denominated in the scarce resource of that detector (machine time for gates; human attention for the outer loop), and let $B_d$ be the sustainable budget of that resource per unit time. The economics of detection impose a simple upper bound on the sustainable feedback frequency:

$$f_d \leq B_d / c_d.$$

Each detector class is bounded in its own resource, so the regimes are compared by affordable frequency rather than by cost. With machine-priced detection, the affordable frequency is high enough to follow the cadence of artifact change: the inner loop. With human-priced detection, it is far lower, too low to follow that cadence, so feedback must be scheduled; this is the outer loop, periodic in the Oregon Experiment's diagnosis [9] and budgeted in Section 4.3. Sustaining machine-rate feedback without exceeding the human-attention budget therefore requires the two detector classes to operate at different feedback cadences; the dual-loop architecture is our realization of that separation (Table 1, P7).